\documentclass[aps,pre,preprint,showpacs,nofootinbib,showkeys,superscriptaddress]{revtex4-1} 
\usepackage[utf8]{inputenc}
\usepackage{natbib}
\usepackage{cmap} 
\usepackage[T1]{fontenc}
\usepackage{microtype}
\usepackage{amsmath,amssymb,mathtools}
\usepackage{txfonts}

\usepackage{mathrsfs}
\usepackage{graphicx,grffile,textcomp,bm}
\usepackage{setspace}
\usepackage{textcomp}
\usepackage{booktabs,tabularx,dcolumn}
\newcolumntype{d}[1]{D{.}{.}{#1}}

\usepackage{url,hyperref}
\usepackage{xcolor}
\usepackage[capitalize]{cleveref}
\crefrangelabelformat{equation}{(#3#1#4)$-$(#5#2#6)}

\usepackage{placeins}
\usepackage[bottom]{footmisc}

\renewcommand\rho\varrho
\renewcommand\vec[1]{\textrm{\bfseries #1}}

\renewcommand\i{\text{i}}

\usepackage[normalem]{ulem}

\begin{document}
\title{Transient coarsening and the motility of optically heated Janus 
colloids in a binary liquid mixture}

\author{Juan Rub\'en Gomez-Solano}
\affiliation{Instituto de Fisica, 
Universidad Nacional Autonoma de M\'exico, Apdo. Postal 20-364, 01000, Ciudad de M\'exico,
M\'exico}

\author{Sutapa Roy}
\affiliation{Max-Planck-Institut f\"{u}r Intelligente Systeme,
   Heisenbergstr.\ 3,
   70569 Stuttgart,
   Germany}
\affiliation{IV. Institut f\"{u}r Theoretische Physik,
   Universit\"{a}t Stuttgart,
   Pfaffenwaldring 57,
   70569 Stuttgart,
   Germany}
 
\author{Takeaki Araki} 
\affiliation{ Department of Physics, 
Kyoto University, 
Kyoto 606-8502, Japan.} 
 
\author{S. Dietrich}
\affiliation{Max-Planck-Institut f\"{u}r Intelligente Systeme,
  Heisenbergstr.\ 3,
  70569 Stuttgart,
  Germany}
\affiliation{IV. Institut f\"{u}r Theoretische Physik,
  Universit\"{a}t Stuttgart,
  Pfaffenwaldring 57,
  70569 Stuttgart,
  Germany}
   
\author{Anna Macio{\l}ek}

\affiliation{Institute of Physical Chemistry, Polish Academy of Sciences, 
Kasprzaka 44/52, PL-01-224 Warsaw, Poland} 
\affiliation{Max-Planck-Institut f\"{u}r Intelligente Systeme,
  Heisenbergstr.\ 3,
  70569 Stuttgart,
  Germany}

\date{\today}

\begin{abstract}
A gold-capped Janus particle suspended in a near-critical binary liquid mixture can 
self-propel under illumination. We have immobilized such a particle in a narrow channel 
and studied the nonequilibrium dynamics of a binary solvent around it, using experiment 
and numerical simulations. For the latter we consider both a purely diffusive and a 
hydrodynamic model. All approaches indicate that the early time dynamics is purely 
diffusive and characterized by composition layers traveling with a constant speed 
from the surface of the colloid into the bulk. Subsequently, hydrodynamic effects set in 
and the transient state is destroyed by strong nonequilibrium concentration fluctuations, 
which arise as a result of the temperature gradient and the vicinity of the critical point 
of the binary liquid mixture. They give rise to a complex, permanently  changing coarsening 
patterns. For a mobile particle, the transient dynamics results in propulsion in the 
direction opposite to that observed after the steady state is attained.

\end{abstract}

\pacs{05.70.Ln, 61.20.Ja, 61.20.Lc, 64.75.+g}
\keywords{Janus colloids, temperature gradient, coarsening, active motion}
\maketitle

\section{Introduction}\label{introduction}
Inspired by biological molecular motors, in recent years there has been an upsurge of 
research efforts to construct artificial devices, which deliver mechanical work or propel 
themselves in a liquid environment. It has been demonstrated experimentally that a 
micron-sized Janus particle, half-coated with metal and suspended in a near critical binary 
solvent, self-propels after illumination with light of low intensity 
\cite{Volpe-et:2011,Buttinioni-et:2012}. Ever since, this type of light activated 
self-propellers has been intensively used for studying active matter. The phenomena, which 
have been explored this way, range from clustering and phase separation in dense suspensions~\cite{Buttinioni-et:2013}, over
the circular motion ~\cite{Kuemmel-et:2013} and gravitaxis~\cite{Hagen-et:2014} of asymmetric 
self-propelled objects, to phototactic behavior~\cite{Lozano-et:2016,Gomez-Solano-et:2017} 
and self-propulsion in viscoelastic fluids~\cite{Gomez-Solano-et:2016,Lozano-et:2018,Narinder-et:2018,Narinder-et:2019} 
and dense colloidal suspensions~\cite{Lozano-et:2019}. 

Another important direction of research aims at understanding the self-propulsion 
mechanism in these systems, which is rather complex. The source of the active motion 
is provided by the local demixing of a binary solvent, which is observed 
around the Janus colloid after its illumination by light of sufficient intensity, 
such that the metal cap is heated above the lower critical temperature $T_c$ of the solvent. 
Evidence has been gathered that the onset of the motion as well as its direction and its speed 
depend sensitively on system parameters, such as the wetting properties of both 
hemispheres of the Janus particle, the intensity of the illumination, the particle size, 
or the average composition of the solvent. 

Hydrodynamic models shed light on the processes occurring at steady state, i.e., 
after a local demixing has been completed and a droplet, rich in the 
species preferred by the hot hemisphere of the Janus particle, has been formed 
\cite{Gomez-Solano-et:2017,Wuerger2015,samin2016,Araki:2019}. 
However, the relevance of the diffusive dynamics, which dominates the 
coarsening process at early and transient times, is still unclear. So far, the early stage 
diffusive dynamics of a local demixing near a Janus particle has been studied only for 
temperature quenches, which do not cross the binodal of demixing~\cite{Roy-et:2018}. 
Deeper quenches have been studied only for homogeneously heated particles~\cite{Roy-et:2018a}.

This has motivated us to combine experiment and theory in order to investigate the
non-equilibrium coarsening dynamics around a hot spherical Janus particle suspended in a 
binary solvent. This is a challenging task because of technical limitations of both approaches. 
Concerning the experiments, one has to use large particles in order to be able to 
resolve the coarsening patterns. On the other hand, simulations for 
large particles are restricted by the availability of computing power. 
Accordingly, a quantitatively reliable comparison of the corresponding 
results is not always possible. We used an optical microscope to measure the time evolution of 
the surface layers and the subsequent growth of a droplet near the heated golden hemisphere. 
After implementing a suitable confinement, a single colloidal particle can be immobilized, which allows one 
to perform accurate measurements of the composition profiles around the particle as a function 
of time. In order to be able to judge the relevance of hydrodynamic effects 
in the transient dynamics, we compare the experimental data with the results 
obtained from two models, i.e., purely diffusive model B and model H which includes hydrodynamics.   

We find that the coarsening dynamics around immobilized Janus particles 
is more involved than the one observed for a self-propelling particle: the coarsening patterns 
which form around a fixed particle at later times are not observed around a mobile particle. 
Such patterns might, however, be relevant for Janus particles in optical potentials, 
which have been recently studied in the context of, e.g., clustering~\cite{Volpe-et:2019} or of
microscopic engines powered by the local demixing of a critical binary liquid 
mixture~\cite{Volpe-et:2018}. The transient dynamics has consequences for the self-propulsion. 
We observe that the Janus particle starts to move long before the stationary-state droplet is 
formed near its hot hemisphere. Interestingly, in these transient states the direction of 
self-propulsion is opposite to the one in the stationary state.

\section{Experiment}\label{exp}

\subsection{Experimental Setup}\label{expS}

\begin{figure}[]
\includegraphics[width=0.8\textwidth]{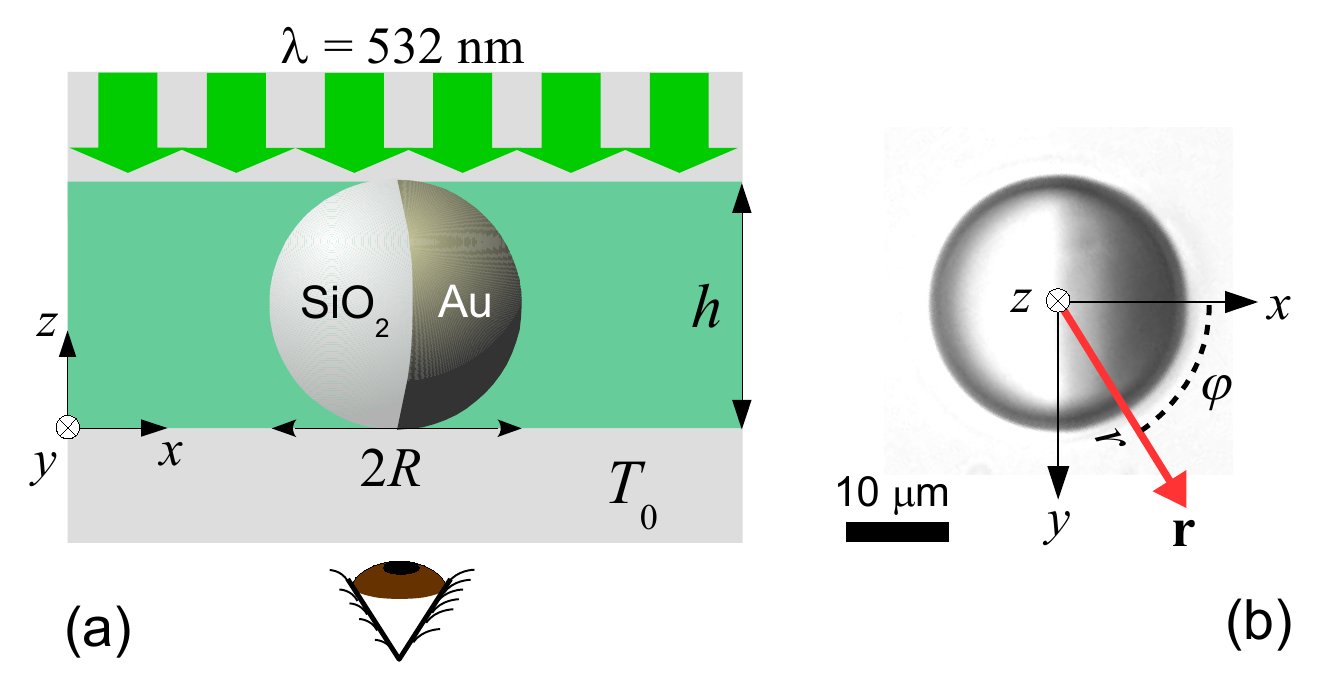}
\caption{(a) Schematic representation of the side-view ($x$-$z$ plane) of 
the experimental setup. The coarsening of the binary liquid mixture (green) 
around the colloid is recorded from below (see the eye) the  sample cell at a temperature $T_0$. 
The sample cell is composed of two glass plates (grey). 
The thick green arrows indicate the illumination by green laser light with  wavelength $\lambda =$ 532nm. (b) The camera image of a Janus colloid with the 
coordinates $\vec r=(x=r\cos\varphi,y=r\sin\varphi)$ is used to describe the coarsening dynamics 
of the surrounding binary liquid mixture in the $x$-$y$ plane. The dark area on the right side of the colloid
corresponds to the golden cap. The origin of the coordinate system is at the point where the spherical colloid touches the surface of the bottom glass plate. For presentation purposes in Fig. 1(a) the coordinate system is shifted to the left. Figure 1(b) corresponds to the projection onto the colloid to the plane $z=0$ (see Fig. 1(a)). The polar coordinates $\varphi$ and $r$ of $\vec r$ within the plane $z=0$ are indicated in (b). The red vector $\vec r$ with $\mid \vec r\mid =r$ is the projection of the three-dimensional vector onto the plane  $z=0$.   }
\label{fig:fig1}
\end{figure}

We use spherical colloids (radius $R=11.6 \pm 0.4 \,\mu\mathrm{m}$) made of silica and 
half-coated by thermal evaporation with a layer of gold (thickness 20~nm). 
The gold caps are chemically functionalized with either 11-mercaptoundecanoic acid 
dissolved in ethanol or 1-octadecanethiol dissolved in ethanol to make them 
strongly hydrophilic or hydrophobic, respectively, while the uncapped silica hemispheres 
remain hydrophilic in all our experiments. 
The particles are suspended in a binary liquid mixture of propylene glycol n-propyl ether (PnP) 
and water, the lower critical point of which is $T_c = 31.9^\circ$C and 
0.4~PnP~mass~fraction~\cite{Bauduin-et:2004}. The binary solvent is at its critical composition. 
At such a concentration, by quickly increasing the temperature from below to above $T_c$, the 
bulk of this binary fluid demixes by spinodal decomposition.
A very small volume ($\sim \mu\mathrm{l}$) of the dilute colloidal suspension is kept within a 
thin sample cell composed of two parallel glass plates, the separation of which 
($h \approx 2 R$) is fixed by the diameter of the particle, as shown in Fig.~\ref{fig:fig1}(a). 
Under such a strong confinement, the largest particles ($R=12 \,\mu\mathrm{m}$) are 
totally immobile while the smallest ones ($R=11.2 \,\mu\mathrm{m}$) experience a large hydrodynamic friction. 
The lateral lengths of the cell along $x$ and $y$  are both 1 cm, i.e., approximately 400 
times the particle diameter. The cell is kept at $T_0 = 28.0^{\circ}$C, in order to maintain 
the bulk fluid in the mixed phase.  The colloid of interest  is chosen in such a way that both 
capped and uncapped hemispheres are equally visible from below the sample cell, i.e. in the 
$x$-$y$ plane, as shown in Figs.~\ref{fig:fig1}(a)-(d). 
Then, an inverse temperature quench is induced around the colloid at time $t=0$ by perpendicularly 
applying green laser illumination ($\lambda = 532$~nm) onto the sample in the $z$ direction, 
as sketched in Fig.~\ref{fig:fig1}(a). Due to the absorption peak of gold around $\lambda = 532$~nm 
and the poor absorption of silica and of the surrounding fluid at that wavelength, the temperature 
non-isotropically increases around the particle surface. For the applied laser intensity 
($\approx 2\,\mu\mathrm{W}\,\mu\mathrm{m}^{-2}$), the final temperature of the cap is 
$T_0 + \Delta T = 35.8^{\circ}$C, which is $3.9^{\circ}$C above $T_c$, thus resulting in a 
local demixing of the binary fluid around the colloid. 

We investigate the coarsening dynamics of the fluid at $t \ge 0$ until its final 
steady-state temperature and concentration profiles are attained. 
For this purpose, using a collimated beam from a second light source which illuminates the sample cell from above ($z \gg h$), in combination with a CCD camera located below ($z < 0$), we record images (in the $x$-$y$ plane) of the light which traverses the fluid around a single colloid and arrives at the CCD sensor
with a sampling frequency of 150 frames per second and a spatial resolution of $0.080\,\mu\mathrm{m}$ per pixel. 
The relevant coordinates ${\bf{r}} = (x = r \cos \varphi, y = r \sin \varphi)$ to describe the coarsening dynamics of the binary fluid are illustrated in Fig.~\ref{fig:fig1}(b), where $r$ is the radial
distance from the south pole of the colloid and $\varphi$ is the azimuthal angle with respect to the $x$ axis pointing
from the uncapped to the capped hemisphere. The third coordinate $z$ of the three-dimensional space does not appear here because the imaging lens, which is positioned beneath the bottom of the sample, provides an image integrated over the vertical coordinate $z$. 

Due to the difference between the refractive index of water and PnP within the temperature range of the experiments ($30\pm 5^{\circ}\mathrm{C}$), $n = 1.331 \pm 0.001$ and $n =1.410 \pm 0.002$, respectively, 
we can perform shadowgraph visualization \cite{steinberg1989,assenheimer1994,mauger2012} of the phases rich in each component during the coarsening process. 
In the mixed phase at  $T_0 < T_c$, the refractive index of the quiescent binary liquid is homogeneous, i.e., $n(r,\varphi,z) = const.$ for $r > R$ and $0 < z < h$, thus leading to a constant light intensity $I_0$ arriving at the camera, i.e., $I(r,\varphi) = I_0$ for $r > R$. After the temperature quench, spatial variations of the refractive index due to local disturbances in the concentration field of the fluid give rise to deviations $\Delta I(r,\varphi) = I(r,\varphi) - I_0$ of the light intensity $I(r,\varphi)$ from the unperturbed intensity $I_0$ \cite{mauger2012}:
\begin{equation}
\label{eq:ex1}
    \Delta I (r,\varphi) \propto  - {I_0} \int_0^{h} \nabla^2 \ln n(r,\varphi,z) \,dz, 
\end{equation}
where $\nabla^2$ is the two-dimensional Laplacian operator in the $x$-$y$ plane. Therefore, this technique allows us to identify bright regions (with respect to $I_0$) in the recorded image, i.e., $\Delta I(r,\varphi) > 0$, as PnP-rich layers, whereas water-rich layers correspond to dark regions, for which  $\Delta I(r,\varphi) < 0$. 
In general, the refractive index of a binary liquid mixture depends on the concentration  in the nonlinear fashion\footnote{However, for a  water-2,6-lutidine mixture it was demonstrated experimentally~\cite{Beysens} that close to the lower critical point of demixing the relation between the refractive index and the concentration of the 2,6-lutidine mixture can be well approximated by a linear function.
We are not aware of similar studies for the mixture of water and PnP used in the present experiment.}.

\subsection{\label{res-expt}Experimental Results}
Depending on the wetting properties of the gold hemisphere, two distinct coarsening patterns can develop. 
This is demonstrated in Figs.~\ref{fig:fig2} and~\ref{fig:fig3} for a hydrophilic and 
a hydrophobic cap, respectively, where, after the quench, we show various stages 
 of the image intensity $\phi := \text{const} \times \Delta I$ (arbitrary units).
We assign $\phi = 0.5$  to the fully mixed fluid while $\phi > 0.5$ and $\phi < 0.5$ represent 
locally PnP-rich and water-rich regions, respectively. We have estimated the resolution $\delta \phi$ 
of the image intensity $\phi$ from its standard deviation for an image of 
the fully mixed binary liquid mixture. Therefore, any variation of $\phi$ less than $\delta \phi$ 
is pure noise caused by the fluctuations of the intensity on the colormap. 
The value we found is $\delta \phi = 0.03$.

We have checked that before and just at the inverse temperature quench, 
i.e., for $t \le 0$, one has $\phi = 0.5$ constant in space and time, as shown in Figs.~\ref{fig:fig2}(a) and ~\ref{fig:fig3}(a). 
In order to characterize the coarsening dynamics, we determine the radial profile of $\phi$ along the 
main particle axis, \i.e., for $\varphi =0$, as a function of the distance from 
the capped particle surface, normalized by the particle radius: 
$\rho = \frac{r-R}{R}$. Certain examples of such radial profiles are plotted in 
the insets of Figs.~\ref{fig:fig2} and~\ref{fig:fig3}.

\subsubsection*{\label{hydrophilic}B.1 Hydrophilic cap}

\begin{figure}[]
\includegraphics[width=\textwidth]{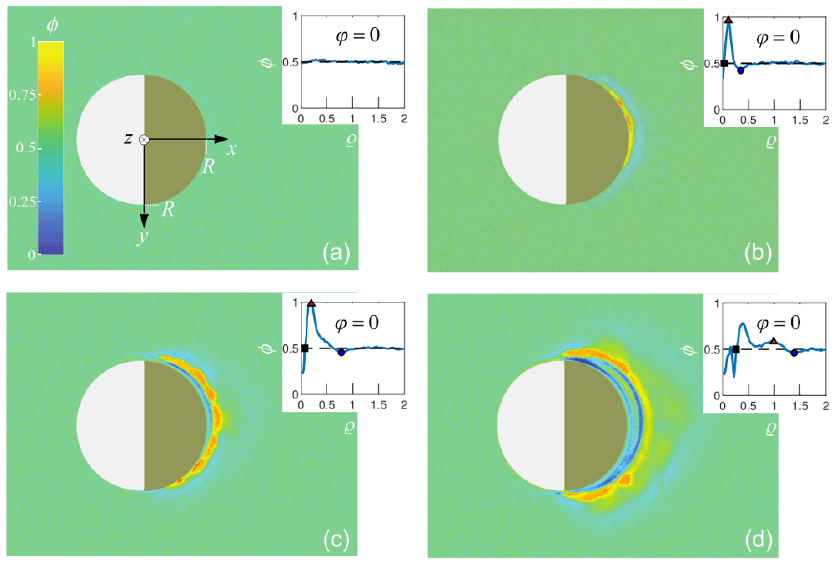}
\caption{Experimental image intensity profile (arbitrary units) during the coarsening of the binary liquid mixture 
around a Janus colloid with a \textit{hydrophilic} cap (right)  at four times after the quench: (a) $t = 0$, 
(b) $t = 0.2$~s, (c) $t=0.4$~s, and (d) $t = 0.8$~s. Here $\phi = 0.5$ corresponds to the fully mixed fluid while 
$\phi > 0.5$ and $\phi < 0.5$ represent 
locally PnP-rich and water-rich regions, respectively. Insets: corresponding radial profiles $\phi(\rho=(R-r)/R)$ for $\varphi=0$. 
The symbols represent the location of the transient PnP layer (red triangle), 
the water droplet thickness (black square) and the location of water rich layer (blue circle).
}
\label{fig:fig2}
\end{figure}

For a hydrophilic cap, we first observe that around the cap a transient water-rich layer immediately forms 
after the quench, as shown in Figs.~\ref{fig:fig2}(b)-(d). This layer radially 
moves away from the particle surface, until it eventually vanishes at sufficiently long times 
($t \approx 1.2$~s). The radial position $\rho$ of this layer is determined by locating the minimum of 
$\phi < 0.5$ as function of time, as illustrated in the insets of Figs.~\ref{fig:fig2}(b)-(d) (circles). 
In Fig.~\ref{fig:fig4}(a) we show that the position of the layer evolves linearly as function of time, 
i.e., $\rho \propto t$, thus moving at constant speed ($\approx 20 \, \mu\mathrm{m}\, \mathrm{s}^{-1}$). 
Moreover, this layer is followed by the formation of a PnP-rich layer, as represented in 
Figs.~\ref{fig:fig2}(b)-(d), the position of which is determined by 
finding the maximum of $\phi > 0.5$ (see the triangles in the insets of Figs.~\ref{fig:fig2}(b)-(d)). 
Unlike the outer water-rich layer, this second layer moves in a nonlinear fashion. While at short 
times it accelerates, it eventually slows down, until it fully disappears at $\approx 1.5$~s 
(see  Fig.~\ref{fig:fig4}(a)). In addition, a droplet rich in water ( $\phi < 0.5$) develops around the 
particle surface ($\rho = 0$), as shown in Figs.~\ref{fig:fig2}(b)-(d). 
Its thickness, inferred from the  location closest to the particle surface at which $\phi = 0.5$ 
(see the squares in the insets of Figs.~\ref{fig:fig2}(b)-(d)), increases nonlinearly in time. 
But unlike the two transient layers, it becomes stable and reaches a finite steady-state size after 
$t \approx 2.5$~s. Indeed, in  Fig.~\ref{fig:fig4}(a) we show that, while the droplet initially 
exhibits a quadratic growth, i.e.,  $\rho \propto t^2$, it gradually levels off with a final thickness 
which is 60\% the particle radius.

\subsubsection*{\label{hydrophobic}B.2 Hydrophobic cap}
\begin{figure}[]
\includegraphics[width=\textwidth]{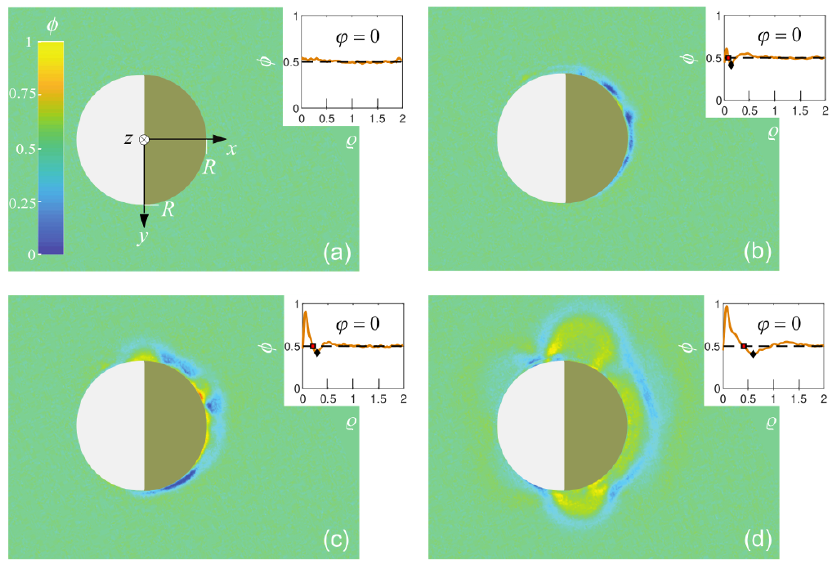}
\caption{Experimental image intensity profile (as in Fig.~\ref{fig:fig2}) during the coarsening of the binary liquid mixture 
around a Janus colloid with a \textit{hydrophobic} cap at four times after the quench: (a) $t = 0$, 
(b) $t = 0.13$~s, (c) $t=0.27$~s, and (d) $t = 0.53$~s; insets: corresponding radial profiles $\phi(\rho=(R-r)/R)$ for $\varphi=0$. The symbols represent the location of the transient water-rich layer (black diamond) and 
the PnP droplet thickness (red square).
}
\label{fig:fig3}
\end{figure}

In the case of coarsening around a hydrophobic cap, we find that a transient water-rich 
layer instantly forms around the cap surface right after the quench, as shown in Figs.~\ref{fig:fig3}(b)-(d). 
The location of such a layer is determined by finding the minimum of $\phi < 0.5$, as depicted 
in the insets of Figs.~\ref{fig:fig3}(b)-(d) (black diamonds).
The water-rich layer propagates away from the cap by enclosing the particle, thereby reaching 
the hydrophilic silica hemisphere.  The layer moves at constant speed 
($\approx 15 \, \mu\mathrm{m}\, \mathrm{s}^{-1}$) and vanishes after $t \approx 0.6$~s, as shown 
in Fig.~\ref{fig:fig4}(b). Then, a stable PnP-rich droplet ($\phi > 0.5$) forms around the cap. 
The thickness of this droplet grows nonlinearly in time. Similar to the profiles around the hydrophilic cap, 
here the droplet thickness is inferred from the location closest to the particle cap at which 
$\phi = 0.5$ (see the red squares in the insets of Figs.~\ref{fig:fig3}(b)-(d)). At the beginning, the 
thickness increases as $\rho \sim t^2$, and then slows down reaching a constant value 
(60\% the particle radius) after $t \approx 1$~s (see  Fig.~\ref{fig:fig4}(b)). We point out that, 
although the coarsening dynamics of the fluid is in this case faster than that for a hydrophilic cap, 
the final thickness of the droplet is the same regardless of the concentration of the dominant component of the 
mixture.  This is the case because the steady-state thickness $\Delta r := r_c - R$ at $\varphi = 0$ of the  
droplet is set by the isotherm $T(r=r_c,\varphi = 0) = T_c$ which serves as an implicit definition of $r_c$. Depending on the wetting properties of 
the cap, a persistent water-rich or PnP-rich phase forms within $R < r \lesssim r_c$, while at $r \gtrsim r_c$ 
the binary fluid remains at the critical concentration.

\section{Simulation Models}\label{simmodel}

In order to understand the observed non-equilibrium dynamics of a binary liquid mixture surrounding 
an immobilized Janus particle after an inverse temperature quench, we adopt two distinct frameworks. 
The first one assumes  purely diffusive dynamics and is the extension of the  Cahn-Hilliard-Cook 
(CHC)-type model, based on the Ginzburg-Landau free energy functional, to non-isothermal systems 
~\cite{Roy-et:2018,Roy-et:2018a}.
This approach was used to study coarsening of the solvent structure surrounding a homogeneous colloidal 
particle, which emerges after a temperature quench at the entire colloid surface. 
However, these earlier studies did not take into account heat diffusion through the particle, and 
they considered homogeneous particles. 
The second approach includes hydrodynamics and is based on the ``fluid particle dynamics'' (FPD) method, 
which describes dynamical couplings between the particle, the concentration, and 
the flow field~\cite{PhysRevLett.85.1338,JPhysCondensMatter.20.072101,PhysRevE.73.061506,
SoftMatter.11.3470,SoftMatter.2017.5911,EurophysLett.51.154,EurophysLett.65.214}. 
This method, extended to a non-isothermal situation,  was used to demonstrate that 
the illumination-induced motion of a Janus particle in binary solvents is not due to  diffusiophoresis but rather due to  Marangoni-like effects~\cite{Araki:2019}.

\subsection{Diffusive dynamics}
\subsubsection*{A.1 Basic equations and boundary conditions}
\label{subsec:DiffusiveD}

Within this approach, the time evolution of the reduced temperature 
$\mathcal{T}(\vec r,t) ~ = \mathcal{A}  (T(\vec r,t)-T_c)/T_c$ and of the concentration field $\psi(\vec r,t)$ 
around the Janus particle, after the temperature quench of the cap, is described by the modified CHC
equation, concertedly with the heat diffusion equation:
\begin{equation}\label{CHC:1} 
\frac{\partial \psi}{\partial t} = \nabla ^2 \Big (-\frac{\mathcal{T}}{|\mathcal{T}_1|} \psi + 
\psi^3 - C\nabla^2 \psi \Big)+ \zeta
\end{equation}
and
\begin{equation}\label{heat1} 
\frac{\partial \mathcal{T}}{\partial t} = {\mathcal D}\nabla^2 \mathcal{T}-s\mathcal{T}.
\end{equation}
These equations are put into dimensionless form by applying a suitable rescaling (see Appendix~\ref{app:CHC}). 
We have assumed a lower critical point, i.e.,  $\mathcal{T}(\vec r,t) < 0$ for the mixed phase.
$\mathcal{T}_1$ is the reduced inverse quench temperature of the cap. 
The magnitude of the sink term $s$ in Eq.~(\ref{heat1}) controls the location of the critical isotherm around the colloid. 
The Gaussian white noise $\zeta$ obeys the relation 
$\langle \zeta(\vec r,t) ~\zeta(\vec r', t')\rangle= -2\zeta_0(\vec r) \nabla^2 \delta (\vec r-\vec r') \delta (t -t')$; 
$\zeta_0(\vec r)$ is the strength of the noise.
Here, $\mathcal D|\mathcal{T}_1|=D_{th}/D_m$ is the Lewis number~\cite{Lewis},
which is the ratio of the thermal diffusivity $D_{th}$ 
and the mutual diffusivity $D_m$ of the solvent.

We employ a no-concentration-flux boundary condition (b.c.) on the surface of the colloid. 
The selective surface adsorption on the colloid surface gives rise to the static so--called Robin b.c. 
~\cite{diehl1992,diehl1997}
\begin{equation}\label{bc2} 
({\hat e_{\mathscr S}} \cdot \nabla \psi(\vec r) + \alpha_{s} \psi(\vec r))|_{{\mathscr S}}=h_s
\end{equation}
where ${{\mathscr S}}$ refers to the surface of the colloid, and $\hat e_{\mathscr S}$ 
is the unit vector perpendicular to $\mathscr S$ pointing into it. 
We assume that the surface enrichment parameters are $\alpha_{s,c}$ ($\alpha_{s,l}$) and that 
the symmetry breaking surface fields are $h_{s,c}$ ($h_{s,l}$) on the cap (on the left hemisphere) of the colloid. 
All surface parameters are made dimensionless (see Appendix~\ref{app:CHC}). 
The Janus colloid is kept confined between two identical parallel walls. We apply the surface b.c. (\cref{bc2}) 
to these walls with the surface parameters $\alpha_{s,w}$ and $h_{s,w}$. 
The cap of the colloid is always maintained at the inverse quench temperature $ \mathcal{T}_1$:
\begin{eqnarray}\label{bc4} 
 \mathcal{T}(\vec r)|_{\mathscr S_c} &=&  \mathcal{T}_1,
\end{eqnarray}
where $\mathscr S_c$ stands for the surface of the cap.
In order to consider the heat flow across the colloid (from the cap towards the left surface), 
\cref{heat1} without the sink term is solved inside the colloid for 
$\mathcal{D}={\mathcal D}_c$, which differs from that in the solvent. 
A smooth variation of the temperature across the surface of the left hemisphere 
is  ensured by adopting the following b.c.:
\begin{eqnarray}\label{bc5} 
\nabla \mathcal{T}|_{\mathscr S_{l,out}}= \nabla \mathcal{T}|_{\mathscr S_{l,in}}.
\end{eqnarray}
Here, ${\mathscr S_{l,out}}$ and ${\mathscr S_{l,in}}$ refer to the outside and the inside of 
the surface of the left hemisphere, respectively.

\subsubsection*{A.2 Numerical setup}
We keep the spherical colloid of radius $R$ fixed at the centre 
of a rectangular simulation box of side lengths $L_x$, $L_y$, and $L_z$ ($L_z << L_x$ and $L_x~=~L_y$). 
Periodic boundary conditions \cite{allen1987} are applied along the $x$ and $y$ directions whereas the 
confining surfaces are placed at $z=0$ and $z=L_z$. 
The \textit{initial} configuration is generated from a uniform random number distribution such that 
the spatially averaged order parameter (OP)
\begin{equation}
\label{eq:aver_psi}
\bar\psi = \frac{1}{V}\int_V \psi d^3r
\end{equation}
is zero which is the critical value. The average OP  $\bar\psi=0$ is conserved as function of time.  $V$ is the volume available for a binary solvent.
The \textit{initial} temperature throughout the system is set below $T_c$ in a system with 
a lower critical point: $ \mathcal{T}_i(\vec r)= -1$, which together with the value of the amplitude $\mathcal{A}\simeq 46.3$ (see Appendix~\ref{app:CHC}) implies $T_i\simeq 0.98 T_c$.
At $t=0$, the cap is quenched to a temperature 
$ \mathcal{T}_1$ and the subsequent dynamics is studied by simultaneously solving the heat diffusion equation 
for the solvent and for the colloid (as mentioned above). The experimental inverse quench temperature 
$35.8^{\circ}$C translates into the reduced temperature $\mathcal{T}_1\simeq 6$  (which implies $T_1\simeq 1.13T_c$) 
for our simulation model. For numerical purposes, we consider 
$\alpha_s=0.5$ and $h_s=-0.2$ both on the top and the bottom confining surfaces 
and $\alpha_{s,l}=\alpha_{s,c}=0.5$ on both sides of the Janus colloid. 
We set $h_{s,l}=-0.2$ on the left side and  $h_{s,c}=-2$ on the capped side of the  hydrophilic-hydrophilic  colloid.
For the  hydrophilic-hydrophobic particle  we take $h_{s,l}=-0.2$ and $h_{s,c}=2$.
We set $C=4$, $s=0.001$, ${\mathcal D}=100$, and ${\mathcal D}_c=3{\mathcal D}$. 
We assume that the typical molecular size $\mathrm{v}_0$ (see Eq. (A1)) is equal to the amplitude $\xi_0^-=0.1$nm 
of the bulk correlation length (above $T_c$ for the lower critical point). 
The CHC theory units are then $r_0 = 0.35$nm for a length and $t_0=10^{-7}$s for time (see Appendix~\ref{app:CHC}).
The noise amplitude $\zeta_0$ is taken to be uniform in space and equal to $10^{-4}$. Unless mentioned otherwise, $L_x=L_y=120$, $L_z=26$, 
and $R=10$ in  units of $r_0$. Finite-element representation of the spatial derivatives combined with the Euler time integrator 
is used to solve the corresponding equations with the time step $\delta t=0.001$ of integration (in  units of $t_0$). 
In order to implement the b.c. on a spherical colloid a trilinear interpolation method \cite{interpolation} is used.

\subsubsection*{A.3 Results}

We calculate {\it all}  quantities (to be specified later) in each of 26 equally spaced, parallel $x$-$y$ planes lying between 
$z=0$ and $z=L_z=26$ and take the average of them. In the following we call this procedure 'depth-averaging'\footnote{The quantity $A(x,y) = \sum_{j=0}^{26}A_j(x,y)$, where $A_j(x,y)$ is its value at the point $(x,y)$ in the $j$th  $x$-$y$ plane, is called  depth-averaged.   }.

We first present simulation results for the dynamics of the local structure formation 
around a {\it hydrophilic-hydrophilic} colloid ($h_{s,l}=-0.2$, $h_{s,c}=-2$). As in the experiment, the cap is more hydrophilic ($h_{s,c} < h_{s,l}$). 
The  rescaled concentration field  $\tilde\psi(\vec r,t) = (\psi(\vec r,t) +1)/2$ is constructed such that $\psi>0.5$ corresponds 
to the PnP-rich phase and $\psi<0.5$ to the water-rich phase (corresponding to $h_s <0$), as the experimental data do. 
\Cref{fig:fig5} depicts the evolution of the depth-averaged snapshots of $\tilde\psi(\vec r,t)$  at four different times following an inverse thermal quench. 
The right side (brown color) of the colloid refers to the cap, which is subjected to an inverse temperature quench. 
At very early times ($t=0.2$) surface layers form  near the cap. 
With increasing time they propagate into the bulk and become thicker. 
It is interesting to note that at later times ($t=8$) the surface layers enter also the region near the uncapped, 
weakly hydrophilic side of the Janus colloid. This is due to the heat flow across the colloid. 
At late times, the Janus particle becomes covered by a water-rich droplet, which is much thicker on the capped 
side of the colloid.
The condensation of  asymmetric droplets around the Janus colloid was found also for 
temperature quenches which do not cross the binodal~\cite{Roy-et:2018}. 
We note that for such quenches this phenomenon is due to the combination of the Soret and 
surface effects~\cite{Roy-et:2018,Roy-et:2018a}. Thus it involves a different mechanism than that occurring for the 
deep quenches considered here.  

In order to quantify the coarsening patterns shown in \cref{fig:fig5}, 
in \cref{fig:fig6} we plot the rescaled OP profile $\tilde\psi(\vec r,t)$ along the $x$-axis 
as a function of the radial distance $\rho = (r-R)/R$ from the colloid surface. 
On top of the depth averaging these data are averaged over $10$ independent initial configurations and are presented at five times. 
 At very early times a water-rich surface adsorption layer 
(with $\tilde \psi<0.5$) forms on the cap, followed by a depletion layer ($\tilde\psi>0.5$) rich in PnP. 
With increasing time, this surface layer thickens and travels away from the colloid surface. 
The maximum of $\tilde \psi$, which corresponds to the depletion layer, also increases and more 
depletion layers form. At later times ($t= 8$ and $50$), the first maximum of $\tilde\psi$ (counting from the colloid surface) decreases, 
both the surface and the depletion layers become very broad, and only one depletion layer is present. 
Far away from the colloid $\tilde\psi(\vec r,t)$ is close to $\tilde\psi_0=0.5$ at all times which means 
that in terms of the radial distance surface layers do not extend. 

Additional information about the formation of the layers can be gained by inspecting 
the vector snapshots of the OP flux which is defined as the spatial gradient of $\psi(\vec r,t)$ normalized to one (see Fig.~1 in the Supplementary Material (SM)).

Next, we present results for a {\it hydrophilic-hydrophobic} colloid ($h_{s,l}=-0.2$, $h_{s,c}=2$).  
All other system parameters are the same as in the case of a hydrophilic-hydrophilic colloid. 
In \cref{fig:fig7}, depth-averaged, evolving snapshots are presented for four different times. 
The color coding is the same as in \cref{fig:fig5}. The capped (right) and uncapped (left) 
hemispheres of the Janus colloid favor the PnP-rich and water-rich phases, respectively, 
which is evident from the color coding in \cref{fig:fig7}. The formation of bicontinuous patterns, as they are characteristic of spinodal decomposition, and of surface layers proceeds in a way similar to \cref{fig:fig5}, i.e.,  at early times only a thin 
surface layer forms and spinodal patterns are more prominent. With increasing time, the thickness of 
this surface layer and of the depletion layer increases  and more layers form. This is readily visible  in \cref{fig:fig8} where  the corresponding rescaled  OP 
profile $\tilde\psi(\vec r,t)$ (depth-averaged and averaged over $10$ independent initial configurations) as a function of $\rho$ is shown at five times.
At late times ($t=100$) only one depletion layer is prevalent. 
In contrast to the case of the hydrophilic-hydrophilic colloid shown in \cref{fig:fig5}, for the hydrophilic-hydrophobic
colloid an asymmetric droplet, condensed at the Janus particle, consists of 
two parts with phases of opposite character: a water-rich part of a droplet, which is thin and 
covers only partly the hydrophilic hemisphere, and a PnP-rich part of a droplet, 
which is thick and extends beyond the capped hemisphere of the Janus particle (see \cref{fig:fig7}). 
The relative size of these parts depends on the surface fields, the initial temperature of the system, 
and the quench temperature.

The temporal evolution of the OP flux around a hydrophilic-hydrophobic
colloid is shown in Fig.~2 in SM. 

From the  OP profiles for a hydrophilic-hydrophilic colloid $\psi(\vec r,t)$ (depth-averaged and averaged over 10 initial configurations) 
we determined the time dependence of the position of the transient water-rich layer by inferring it from 
the second minimum of $\psi(\rho)$, of the position of the transient PnP-rich layer by inferring it 
from the first maximum of $\psi(\rho)$, and of the thickness of the water-rich droplet around the Janus colloid 
by inferring it from the zero crossing of $\psi(\rho)$. 
Similarly, for the hydrophilic-hydrophobic colloid, the time dependence of the position of the 
transient PnP-rich layer is determined by inferring it from the second maximum of $\psi(\rho)$, of the 
position of the transient water-rich layer is determined by inferring it from the first minimum of 
$\psi(\rho)$, and of the thickness of the PnP-rich droplet around the Janus colloid is determined by inferring it 
from the zero crossing of  $\psi(\rho)$. The corresponding results are plotted in \cref{fig:fig10}.

\subsubsection*{A.4 Comparison with experiment}
\begin{figure}[]
\includegraphics[width=0.8\textwidth]{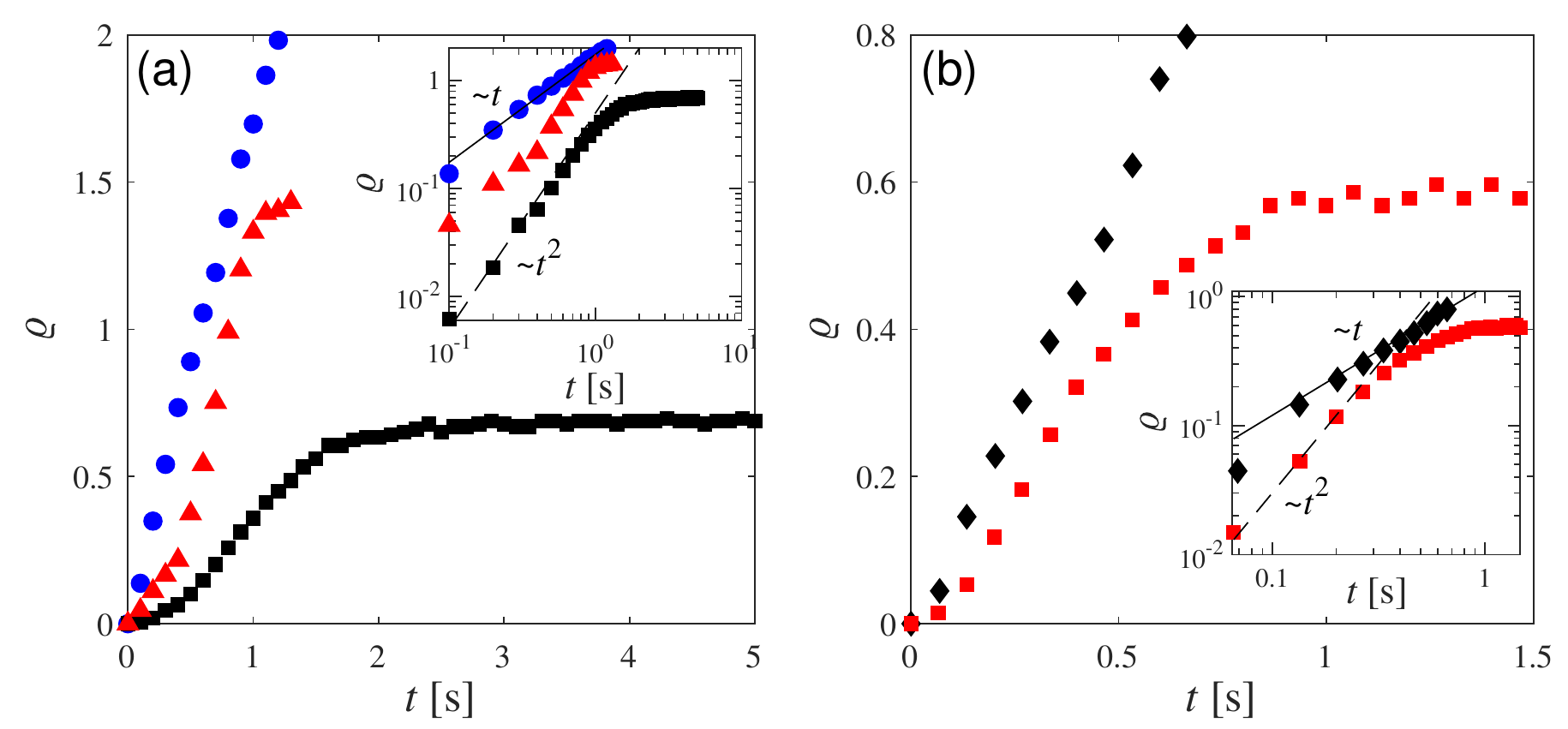}
\caption{(a) Experimental time evolution of the position of the transient water-rich layer (blue circles), 
the position of the transient PnP-rich layer (red triangles), and the thickness of the water-rich 
droplet (black squares) after a quench of the binary liquid mixture surrounding a colloid 
with a \textit{hydrophilic} cap; \textit{inset}: log-log representation of the main figure. 
The lines are guides to the eye in order
to illustrate a linear growth (full line) and a quadratic growth (dashed line), respectively. (b)  Time evolution of 
the position of the transient water-rich layer (black diamonds), and the thickness of the 
PnP-rich droplet (red squares) after a quench of the binary liquid mixture surrounding a 
colloid with a \textit{hydrophobic} cap; inset: log-log representation of the main figure. 
The lines are guides to the eye in order to illustrate a linear growth (full line)
and a quadratic growth (dashed line).}
\label{fig:fig4}
\end{figure}

\begin{figure}[htbp!]
\centering
\includegraphics*[width=0.8\textwidth]{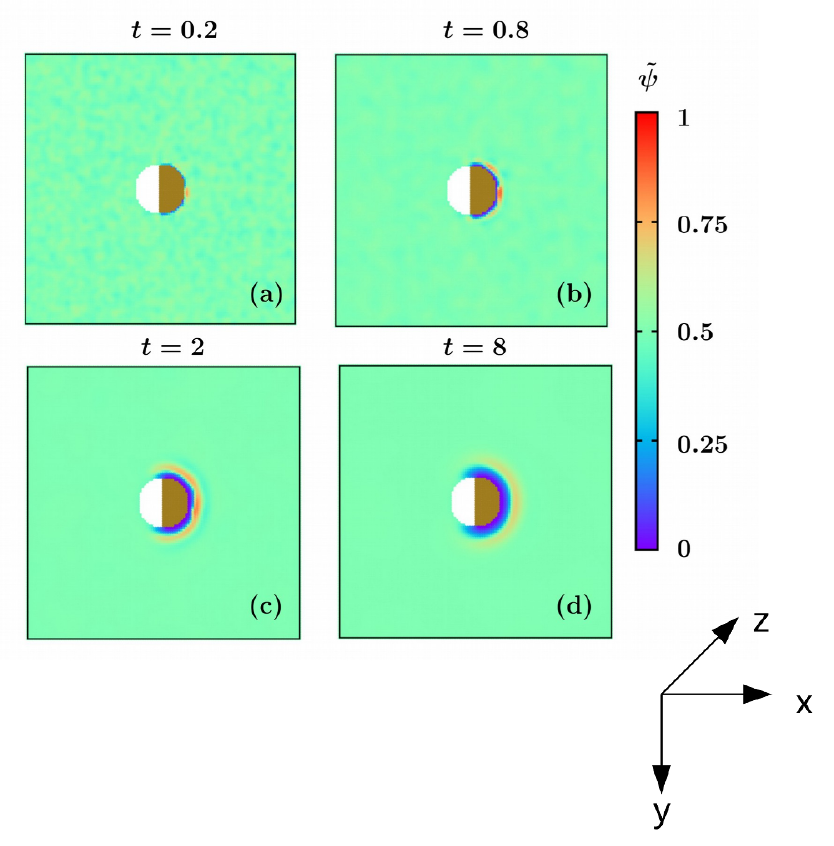}
\caption{Theoretical  evolution snapshots of the rescaled concentration field  $\tilde \psi(\vec r,t) = (\psi(\vec r,t)+1)/2$ during the coarsening of a binary solvent 
under a time-dependent temperature gradient around a \textit{hydrophilic-hydrophilic} 
Janus colloid confined between two parallel slabs kept at $z=0$ and $L_z$. 
In the color code, $\tilde \psi>0.5$ and $<0.5$ refer to the PnP-rich and the water-rich phase, respectively. 
Far away from the colloid one can see  bicontinuous coarsening patterns characteristic of spinodal decomposition. 
Surface layers form due to the surface-directed spinodal-decomposition mechanism. 
With increasing time, these layers broaden and increase in number. The snapshots correspond to the bottom view (Fig. 1(b)) 
and are depth-averaged 
(see the main text). Time is given in units of $t_0=10^{-6}$s. 
The side bar provides the color code for the values of $\tilde\psi$. The brownish color indicates that the right hemisphere is capped. We do not specify  wettablity of the cap via a color code.}
\label{fig:fig5}
\end{figure}

\begin{figure}[]
\centering
\includegraphics*[width=0.8\textwidth]{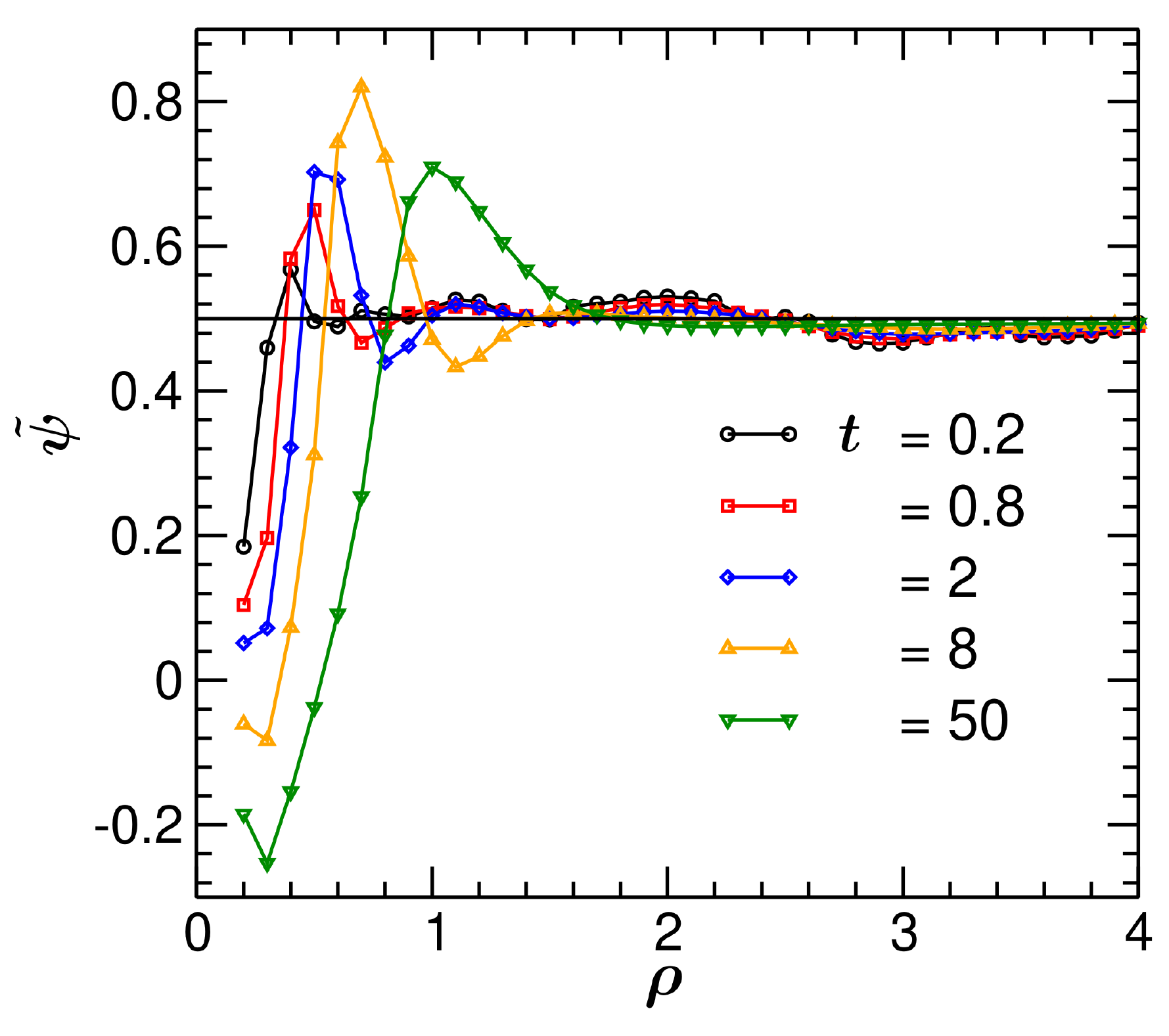}
\caption{Rescaled theoretical OP profiles $\tilde \psi(\vec r,t)$ after depth-averaging and averaged over 10 initial conditions
along the $x$-axis around a \textit{hydrophilic-hydrophilic} Janus colloid as a function of the radial distance $\rho = (r-R)/R$ 
from the surface of the colloid. Time is given in units of $t_0=10^{-6}$s.}
\label{fig:fig6}
\end{figure}

\begin{figure}[]
\centering
\includegraphics*[width=0.8\textwidth]{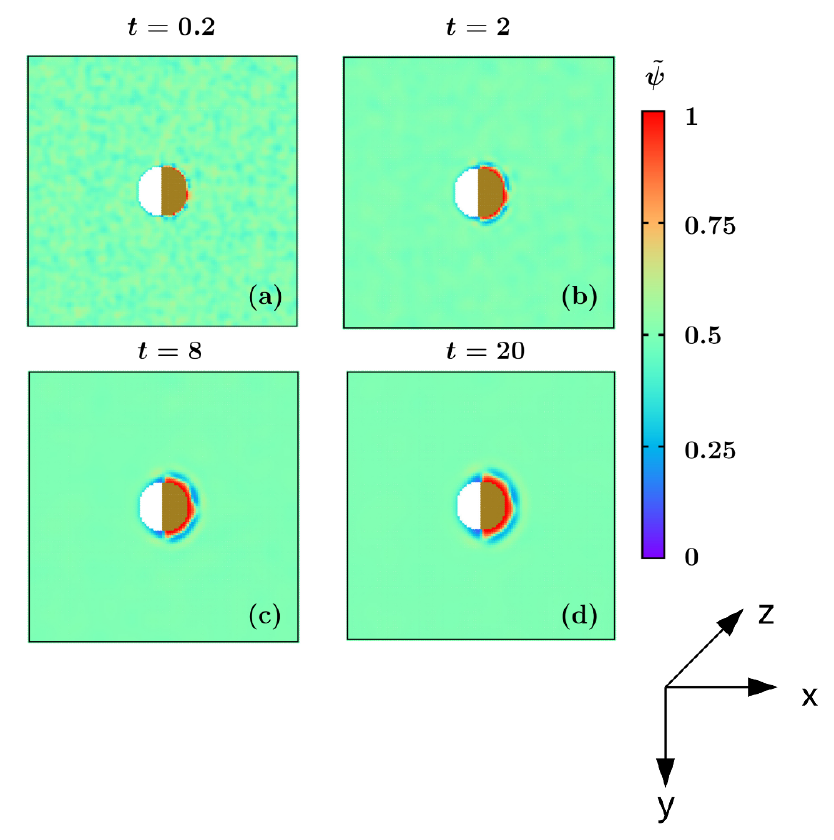}
\caption{Temporal development of the rescaled theoretical concentration field $\tilde \psi(\vec r,t)$  around a 
\textit{hydrophobic-hydrophilic} Janus colloid and its dynamics within the demixing zone. 
In the color code, $\tilde \psi>0.5$ and $<0.5$ refer to the PnP-rich and water-rich phases, respectively. 
The snapshots correspond to the bottom view and are depth-averaged. The brownish color indicates that the right hemisphere is capped. We do not specify  wettablity of the cap via  a color code.
Time is given in units of $t_0=10^{-6}$s.}
\label{fig:fig7}
\end{figure}

\begin{figure}[]
\centering
\includegraphics*[width=0.8\textwidth]{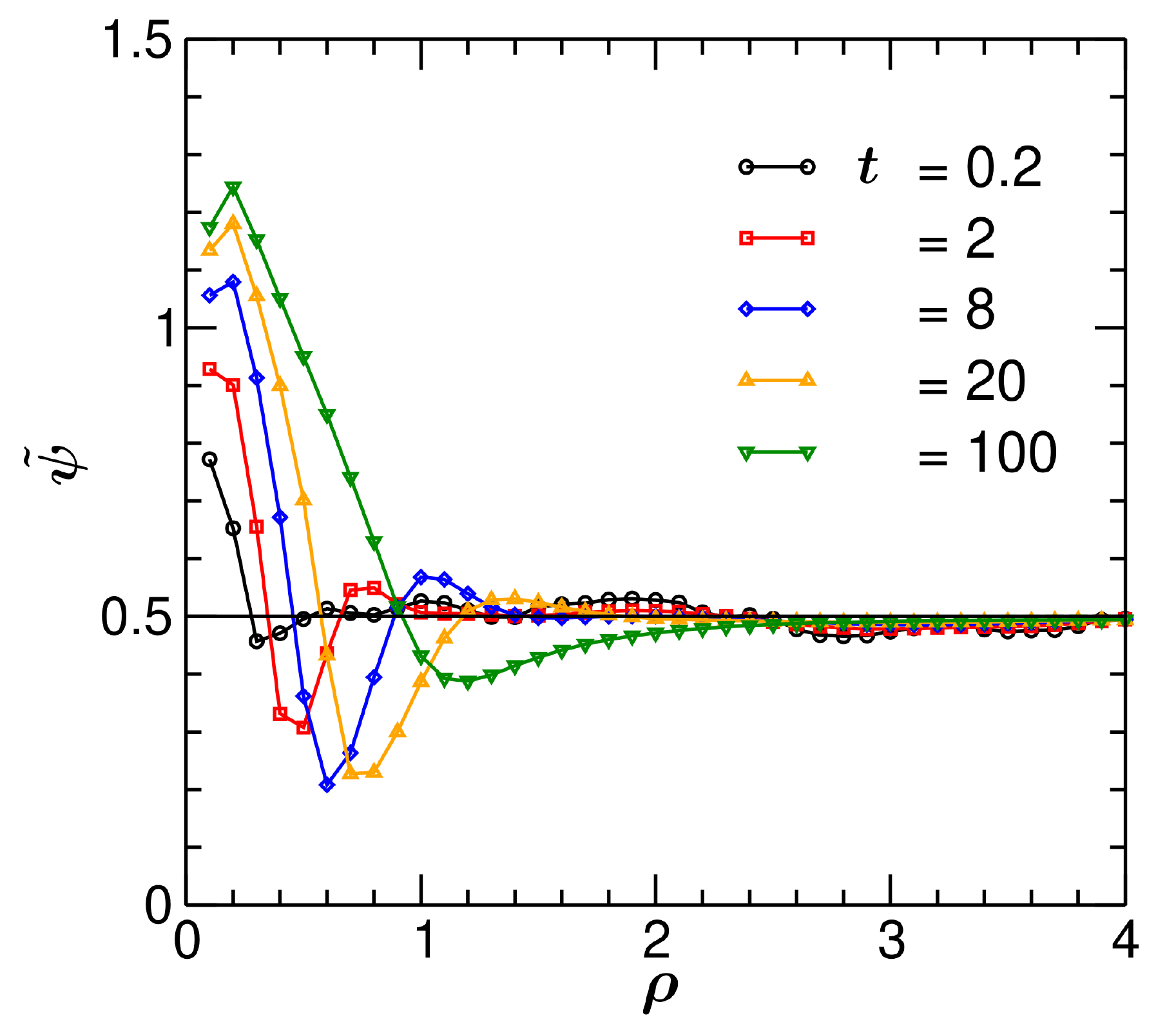}
\caption{Rescaled theoretical OP profiles $\tilde \psi(\vec r,t)=(\psi(\vec r,t)+1)/2$ after a depth averaging and averaged over 10 initial conditions
along the $x$-axis around a \textit{hydrophilic-hydrophobic}  Janus colloid as a function of the radial distance 
$\rho = (r-R)/R$ from the surface of the colloid. Time is given in units of $t_0=10^{-6}$s. }
\label{fig:fig8}
\end{figure}

\begin{figure}
\centering
\includegraphics*[width=0.45\textwidth]{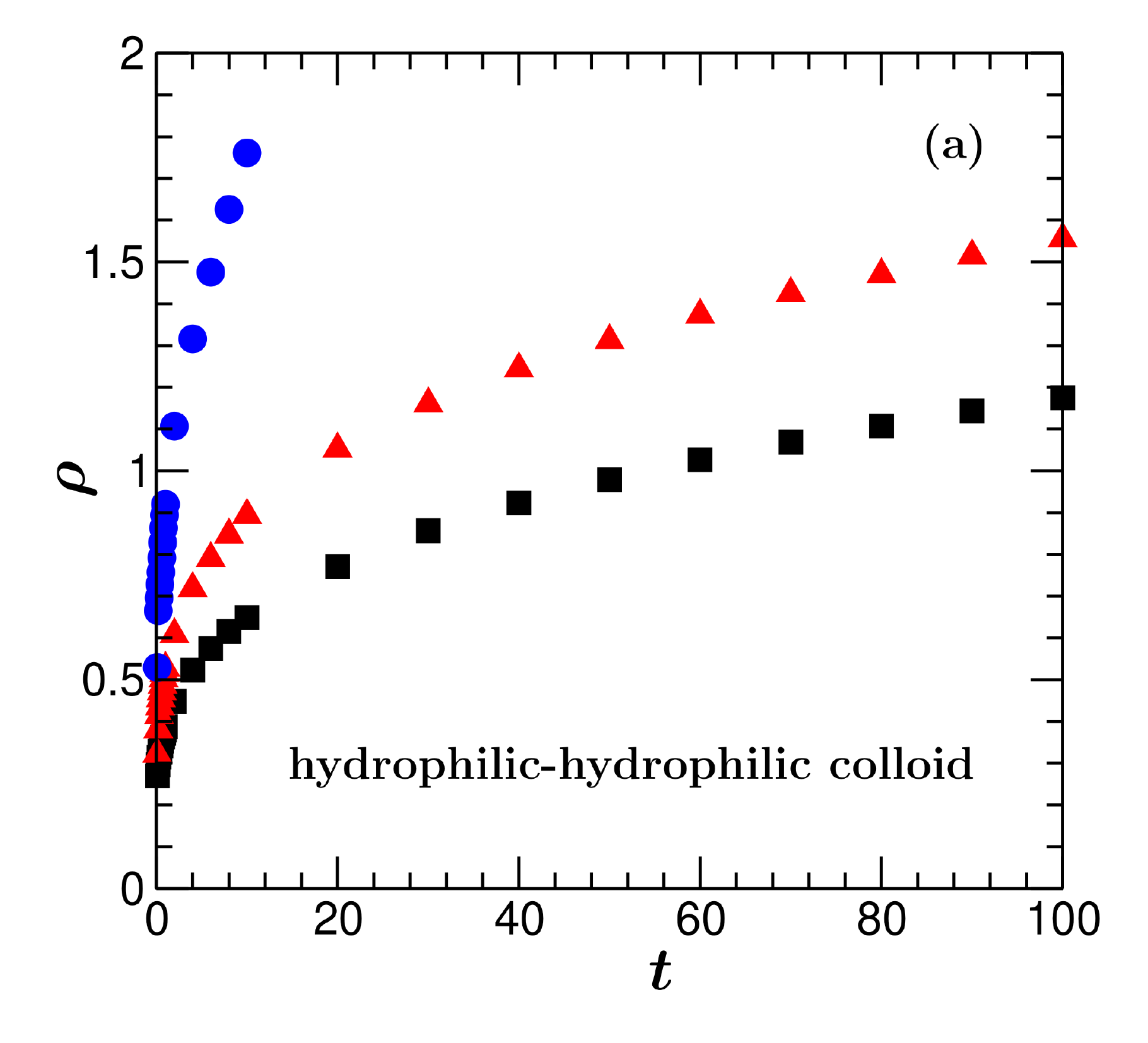}
\includegraphics*[width=0.45\textwidth]{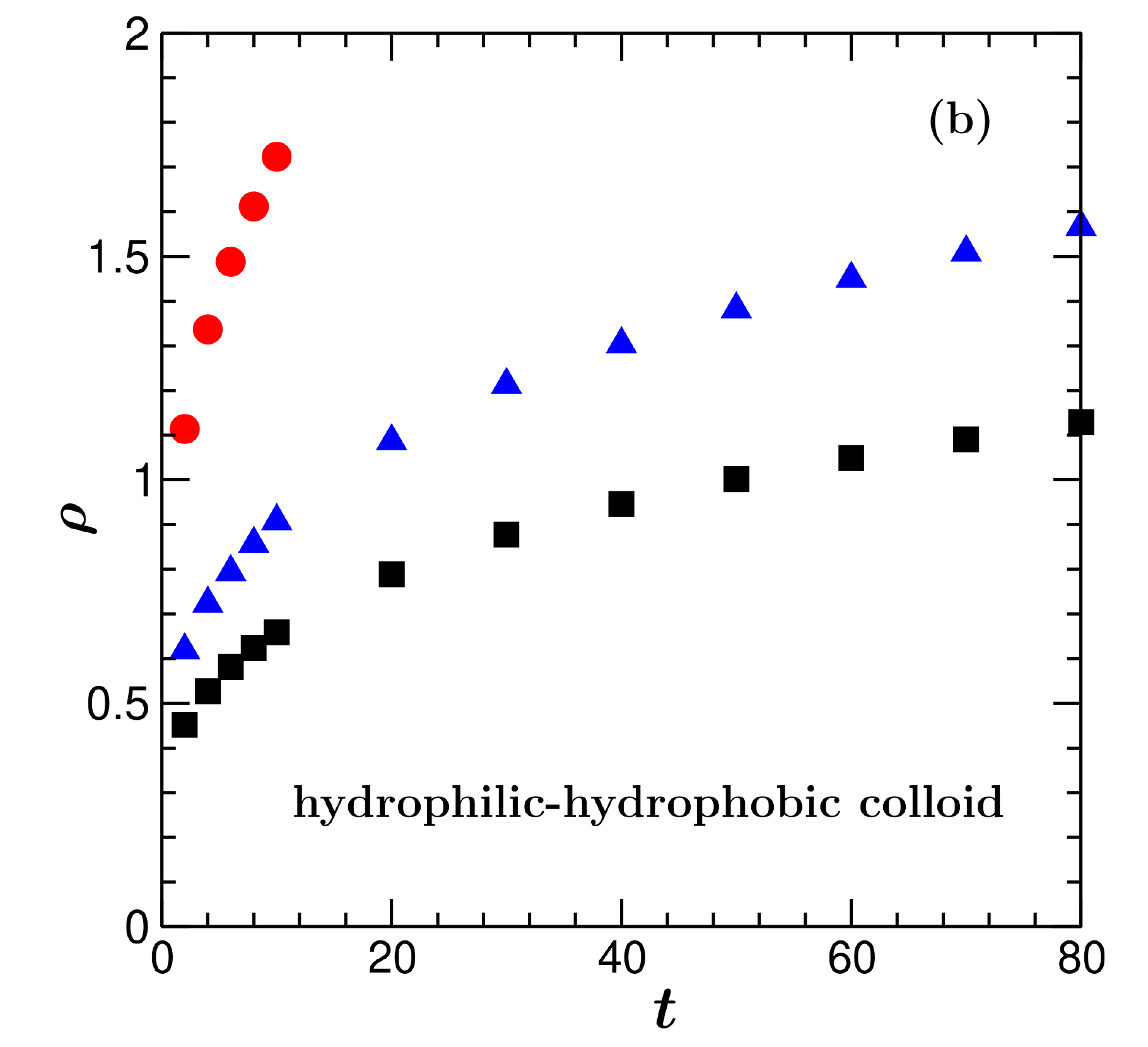}
\caption{Numerical data after depth averaging and averaging over 10 initial conditions from a purely diffusive model concerning (a) the 
time evolution of the position of the transient water-rich layer (blue circles), the position of the
transient PnP-rich layer (red triangles), and the thickness of the water-rich droplet (black squares) 
after a quench of the binary liquid mixture surrounding a \textit{hydrophilic-hydrophilic} colloid, i.e., 
with a hydrophilic cap. (b) Time evolution of the position of the transient PnP-rich layer 
(red circles), the position of the transient water-rich layer (blue triangles), and the
thickness of the PnP-rich droplet (black squares) after a quench of the binary liquid mixture 
surrounding a \textit{hydrophilic-hydrophobic} colloid with a hydrophobic cap.}
\label{fig:fig9}
\end{figure}

As already mentioned in the Introduction, the length and time units in CHC theory are very different from the experimental ones\footnote{In this section we use the subscript ``exp'' in order to distinguish the actual time
from the reduced simulation time.}. Moreover, the choice of the surface interaction parameters is 
to a certain extent arbitrary, because their relation to the materials properties of the
colloid surfaces and  walls is not known. In the theoretical model we have taken the same strength |$h_{s,c}$| of the 
surface parameters on the cap - independent of its sign and thus its wettability, i.e., $|h_{s,c}|=2$ for a  hydrophilic ($h_{s,c}<0$) and a hydrophobic ($h_{s,c}>0$) cap. 
This might not to be the case for actual systems. 
Nevertheless, at short times after the temperature quench we find a qualitatively similar behavior 
of the actual system and the model system. Specifically, snapshots of the image intensity  and the corresponding 
profiles for a particle with a hydrophilic cap at $t_{exp}=0.2$s and  $t_{exp}=0.4$s in Figs.~\ref{fig:fig2}(b) and (c)) 
look similar to the ones for the concentration field computed in simulations for $t=0.8$ and 2 in Figs.~\ref{fig:fig5} and ~\ref{fig:fig6}. 
In the experiment, at the time $t_{exp}=0.8$s, the layer structure around the hot part of the Janus particle 
becomes more diffuse, which is reflected in the measured image intensity  profile. 
This is not found in the simulations in which the layer structure stays sharp even at later times 
(see Fig.~\ref{fig:fig5}(d) and Fig.~\ref{fig:fig6} for $t=50$).
As function of time, the linear increase of the position of the transient water-rich layer as 
observed in the experiment is found also in the simulations for $t\gtrsim 0.8$ after a transient time 
during which strong thermal fluctuations affect the layer formation 
(compare the blue circles in Figs.~\ref{fig:fig4} and \ref{fig:fig9}(a)). 
The constant speed of ca $13\, \mu\mathrm{m}\, \mathrm{s}^{-1}$ at which the transient water-rich layer moves 
is comparable with the experimental result of ca $20 \, \mu\mathrm{m}\, \mathrm{s}^{-1}$. 
As in the experiment, the motion of the  transient PnP-rich layer is nonlinear. 
Concerning the thickness of the water-rich droplet, although the shape of the numerical curve 
(blue squares in Fig. \ref{fig:fig9}(a)) describing its growth in time resembles the experimental one 
(blue squares in Fig. \ref{fig:fig4}(a)), it does not follow the power law $\sim t^2$ 
at early times (inset of Fig. \ref{fig:fig4}(a)).

In the case of coarsening around a hydrophobic cap, there are more discrepancies between the experimental and 
the simulation results.
In the experiment, the time evolution of the transient PnP-rich transient layer could not be measured because, although present, this layer was hardly visible and its position was difficult to infer 
from the second maximum of the radial profile of $\psi$. In the simulations this layer moves with a 
constant speed of ca $11.5\, \mu\mathrm{m}\, \mathrm{s}^{-1}$ for $t\gtrsim 10$. The measurements 
show that the transient water-rich layer moves with a constant speed of ca $15 \mu\mathrm{m}\, \mathrm{s}^{-1}$ 
for $t\gtrsim 10$ (blue diamonds in Fig. \ref{fig:fig4}(b)), whereas in the simulations the motion of the transient water-rich layer  is nonlinear (blue triangles in Fig. \ref{fig:fig9}(b)). 
The thickness of the PnP-rich droplet
(red squares in Fig. \ref{fig:fig9}(b))  does not follow the power law $\sim t^2$ found in the experiment  
at early times (inset of Fig. \ref{fig:fig4}(b)).
In the experiment,  at $t=0.53$ the layer structure is destroyed
and one observes a flower-like structure around the hot cap of the colloid. This points towards a 
different mechanism of the coarsening process, which is not found in the simulations of 
the present, purely diffusive model.

These discrepancies between the experimental observations and the results of the simulations 
of the purely diffusive model could be due to hydrodynamic effects, which will be studied in the next section.

\subsection{Hydrodynamic model}\label{arakimodel}

\subsubsection*{B.1 Formalism}\label{arakimodel:1}

The `` fluid particle dynamics'' (FPD) method \cite{PhysRevLett.85.1338} is a  hybrid  model,  which  combines  a  lattice  simulation  for  continuous  fields describing a binary liquid solvent  and  an  off-lattice  simulation
for an immersed particle. 
It   is  related to the so-called  model H simulations
of fluid phase separation, the hydrodynamics of which   is
also  described  by  the  Navier-Stokes  equation.  The advantage of the FPD method  is, that it avoids problems related to  the discontinuity of the flow fields at the solid-fluid
 boundary, which does occur in the model H simulations. The physical foundations of the FPD method for colloid dynamics simulation are discussed in Ref.~\cite{Furukawa-et:2018}.
 
Within this approach, the Janus particle of radius $R$ is represented  by  a smooth shape function as 
\begin{eqnarray}
\mathcal{S}({\bf r},{\bf x})
&=&\frac{1}{2}\left\{1+
\tanh\left(\frac{R-|{\bf r}-{\bf x}|}{d_{\mathcal{S}}}\right)
\right\}.
\end{eqnarray}
Here, ${\bf x}$ is the position of the center of the particle;
$d_{\mathcal{S}}$ represents the width 
of the smooth interface such that,  
in the limit of $d_{\mathcal{S}} \rightarrow 0$, $\mathcal{S}$ is unity and zero in the 
interior and exterior of the particle, respectively. 
We also define the particle orientation function 
\begin{eqnarray}\label{orien}
\mathcal{O}({\bf r},{\bf x},{\bf n})
&=&\frac{1}{2}\left\{
1+\tanh \left(
\frac{1}{d_{\mathcal{O}}}{\bf n}\cdot \frac{{\bf r}-{\bf x}}{|{\bf r}-{\bf x}|}\right)
\right\},
\end{eqnarray}
where ${\bf n}$ is a unit vector along the symmetry axis of the Janus particle;
$d_\mathcal{O}$ is a sharpness parameter of the particle orientation. 
Roughly, one has $\mathcal{O}=1$ if ${\bf n}\cdot ({\bf r}-{\bf x})/|{\bf r}-{\bf x}|>d_\mathcal{O}$, while $\mathcal{O}=0$ if ${\bf n}\cdot ({\bf r}-{\bf x})/|{\bf r}-{\bf x}|<-d_\mathcal{O}$. In this study, we set $d_\mathcal{O}=0.033$ in order to assure that   the
surface properties change smoothly within a few lattice constants. In the limit of $d_\mathcal{O} \rightarrow 0$, one has
$\mathcal{O}=1$ around the cap (${\bf n}\cdot ({\bf r}-{\bf x})>0$), while $\mathcal{O}=0$ otherwise. 

In order to describe a binary liquid mixture in an inhomogeneous 
temperature field, we employ the dynamic van der Waals 
theory 
\cite{Onuki_book_2002,PhysRevLett.51.054501,PhysRevE.75.036304,
EurophysLett.84.36003,PhysRevE.82.021603,PhysRevE.84.041602} extended 
to incompressible binary liquid mixtures. 
The concentration field of the binary liquid mixture $\psi({\bf r})$ is coupled to the surface of the Janus particle with the
energy $E$  given by 
\begin{eqnarray}
E\{\psi,{\bf x},{\bf n}\}
=\int_{V}d^3 r\left[
\left\{h_{s,l}+(h_{s,c}-h_{s,l})\mathcal{O}({\bf r},{\bf x},{\bf n})\right\}\psi({\bf r})+\frac{1}{2}\alpha_s \psi^2({\bf r})
\right]|d_{\mathcal{S}}\nabla \mathcal{S}({\bf r},{\bf x})|;
\end{eqnarray}
$V$ is the volume of the system, $h_{s,c}$ and $h_{s,l}$ represent the symmetry breaking surface fields on the cap and on the other, i.e., left part of the particle surfaces, respectively, and
$\alpha_s$ is the surface enrichment parameter. 

We assume that the time development of the concentration field $\psi({\bf r},t)$ is governed by 
\begin{eqnarray}
\frac{\partial \psi({\bf r},t)}{\partial t}
&=&-\nabla \cdot \{\psi ({\bf r},t){\bf v}({\bf r},t)\}
\nonumber\\
&&+
\nabla \cdot 
\left\{
(1-\mathcal{S})\nabla
\left(
-\frac{{\mathcal{T}}({\bf r},t)}{|{\mathcal{T}}_1|}\psi({\bf r},t)
+\psi^3({\bf r},t)-C\nabla^2 \psi({\bf r},t)
+\frac{\delta E}{\delta \psi}\right)
\right\}+\zeta({\bf r},t). 
\end{eqnarray}
The first term on the right hand side is the convection term due to the hydrodynamic flow field ${\bf v}({\bf r},t)$. 
The expression $\delta E/\delta \psi=[\{h_{s,l}+(h_{s,c}-h_{s,l}) \mathcal{O}\}+\alpha_s \psi(\mathbf{r})]|d_\mathcal{S}\nabla_{\mathbf{r}} \mathcal{S}|$ represents the wetting interaction on the particle surface. 
The diffusion flux inside the particle vanishes  due to the term  $(1-\mathcal{S})$; $\zeta({\bf r},t)$ is the Gaussian white noise introduced earlier (see Sec.~\ref{subsec:DiffusiveD}).

The time development of the temperature field is given by 
\begin{eqnarray}
\frac{\partial {\mathcal{T}}({\bf r},t)}{\partial t}
&=&-\nabla \cdot ({\mathcal{T}}{\bf v})+\nabla \cdot 
\left[\left\{
\mathcal{D}+(\mathcal{D}_{c}-\mathcal{D})\psi({\bf r},t)
\right\}\nabla {\mathcal{T}}({\bf r},t)
\right]
\nonumber\\
&&+\frac{1}{2}g\{{\mathcal{T}}_1-{\mathcal{T}}({\bf r},t)\}
\mathcal{O}|d_{\mathcal{S}}\nabla_{\mathbf{r}}\mathcal{S}|. 
\end{eqnarray}
$\mathcal{D}_{c}$ and $\mathcal{D}$ are the thermal diffusion constants inside the particle and of the binary solvent, respectively. 
The last term on the right hand side of Eq. (11) is introduced in order to fix the temperature on the cap of the particle. The cap is  represented by $\mathcal{O}|d_{\mathcal{S}}\nabla \mathcal{S}|$, and $g$ is a parameter related to the heating power\footnote{The expression $g(\mathcal{T}_1-\mathcal{T}({\bf r},t))/2$ can be rewritten as $H/C_{\rm p}-s\mathcal{T}({\bf r},t)$, where $H$ and $C_{\rm p}$ are the heating power and the specific
heat (see Ref.~\cite{Araki:2019}). The quantity $-s \mathcal{T}$ represents the dissipation to the bath (see Eq.~(\ref{heat1}) for the purely diffusive model).
In the hydrodynamic model, this dissipation is introduced only at the surface. With fixed temperature at the boundary of the simulation box this controls the location of the critical isotherm around the colloid. The factors $g$ and $\mathcal{T}_1$ are related to $H$, $C_p$, and 
$s$ according to $\mathcal{T}_1=H/(C_p s)$ and 
$g=2s$.}. 

We consider the hydrodynamic flow ${\bf v}({\bf r})$ in the limit of low Reynolds numbers, in which the effect 
of inertia is negligible. 
The flow field ${\bf v}({\bf r})$ is obtained by solving the following differential equations:
\begin{eqnarray}\label{hydr}
&&C\nabla \cdot\left(\nabla \psi:\nabla \psi\right)
-
\frac{\mathcal{S}}{\Omega}
\left(\frac{\partial E}{\partial {\bf x}}\right)
-\frac{1}{2}\nabla \times 
\left\{
\frac{\mathcal{S}}{\Omega}{\bf n}\times
\left(\frac{\partial E}{\partial {\bf n}}\right)
\right\}
\nonumber\\
&&
-\nabla p
+\nabla \cdot\left[
\{\eta+(\eta_c-\eta) \mathcal{S}\}\
\left\{
\nabla: {\bf v}+(\nabla: {\bf v})^T
\right\}\right]+K({\bf x}-{\bf x}_0)\frac{\mathcal{S}}{\Omega}
=0. 
\end{eqnarray}
The first term is the mechanical stress stemming from the concentration inhomogeneity. 
The second and third terms are due to the coupling energy $E;$
$p$ is the pressure  obtained via the incompressibility condition $\nabla \cdot {\bf v}=0$. 
Within the FPD scheme the fifth term is due to the viscous stress, in which $\eta$ and $\eta_c$ are the viscosity of the solvent and inside the particle, respectively. 
The last term is introduced in order to fix the particle at its initial position ${\bf x}_0$ by imposing a harmonic potential with  spring constant $K$. 

The particle position and its orientation are transported by the hydrodynamic flow ${\bf v}$ and its vorticity according to
\begin{equation}
\frac{d}{dt}{\bf x} = \frac{1}{\Omega}\int_{V}d^3r\left\{\mathcal{S}({\bf r},{\bf x}){\bf v}({\bf r})\right\}
\end{equation}
and
\begin{equation}
\frac{d}{dt}{\bf n} = \frac{1}{2\Omega}\int_{V}d^3r\left\{\mathcal{S}({\bf r},{\bf x}){\bf n}\times (\nabla \times 
{\bf v}({\bf r}))\right\},
\end{equation}
where $\Omega =\int_{V}d^3r \mathcal{S}({\bf r},{\bf x})$ is the particle volume.

\subsubsection*{B.2 Numerical setup}\label{arakimodel:2}
For numerical purposes we assume that   ${\bf r}/d_{\psi}\in \mathbb{Z}^3$ are the simple cubic lattice coordinates in lattice space, whereas the position ${\bf x}$ of the center of the particle is in the off-lattice space, i.e., the components of ${\bf x}$ are floating point numbers. 
The time increment is $\delta t=0.005$. 
The system size is $240\times 240\times 52$. 
Periodic boundary conditions are employed in the $x$ and $y$ directions. 
On the other hand, we employ  non-slip boundary conditions for the flow field on the bottom ($z=1$) and  top ($z=52$) planes. 
The concentration fluxes vanish on these walls, too. 
The temperature at these walls is fixed to $\mathcal{T}_i=-1$ 
and after the quench that on the cap is controlled to  be $\mathcal{T}_1=6$ with $g=10$. 
The thermal diffusion constants are $\mathcal{D}=100$ and $\mathcal{D}_c=3\mathcal{D}$. 
These walls act as a heat reservoir. 
The particle radius is $R=20$.
Initially, the particle is placed at the center of the cell and is oriented in the $(1,0,0)$-direction. 
In the present study, the particle position and orientation are almost fixed at their initial values owing to the harmonic potential $K({\bf x}-{\bf x}_0)^2/2$. 
The parameters in the hydrodynamic equation (\ref{hydr}) are taken to be $C=4$, $K=100$, $\eta=0.5$, and $\eta_s=25$. For the length unit  $d_{\mathcal{S}}$ (Eq.~(\ref{orien})) we take the typical value 2nm. 
The time unit is given by $t_0=d_{\mathcal{S}}^2/D_m$. Assuming $D_m \simeq 1.1\times 10^{-10}$m$^2$s$^{-1}$ gives $t_0\simeq 3.6\times 10^{-8}$s. 
The amplitude $\zeta_0$ of the noise  is the same as in the purely diffusive model, i.e., $10^{-4}$.

In order to mimic an  experimental system, we consider two kinds of Janus particles with different surface fields. 
One particle has a strongly hydrophilic cap and a weakly hydrophilic tail, for which we set $h_{s,c}=-2.0$ and $h_{s,l}=-0.2$. 
The other particle has a strongly hydrophobic cap and a weakly hydrophilic tail with $h_{s,c}=2.0$ and $h_{s,l}=-0.2$. 
The top and bottom walls of the cell are weakly hydrophilic, for which we set $h_{wall}=-0.2$. 
The surface enrichment parameter is $\alpha_s=0.5$. 
We note that the choice of the surface interaction parameter is the same as in the diffusive model.

For the initial configuration  we assume that 
the concentration field is $\psi=0$ throughout and that the temperature is
 $T=-1$ everywhere. The flow is absent.
The particle is placed at the center of the cell and it is oriented
along the $x$-axis. The particle position and orientation are almost
fixed during the simulations.

\subsubsection*{B.3 Results}\label{arakimodel:3}
\begin{figure}[thbp!]
\centering
\includegraphics*[width=0.8\textwidth]{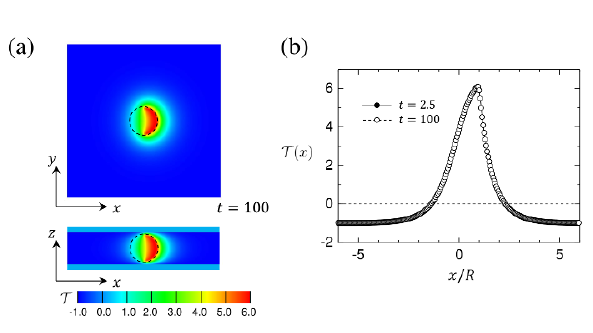}
\caption{(a) 
Snapshot of the temperature field around a \textit{hydrophilic-hydrophilic} Janus particle in the $x$-$y$ and $x$-$z$ planes passing through the particle center of the system. 
The temperature at the bottom and the top walls of the cell is fixed at $\mathcal{T}_i=-1$ and the  temperature of  the surface of the capped hemisphere of the particle is quenched to 
 $\mathcal{T}_1=6$.
The average concentration (see Eq.~(\ref{eq:aver_psi})) is $\bar{\psi}= 0$ and conserved as function of time. 
The black circle represents the particle, which is oriented 
to the right, i.e., the heated  particle cap is oriented towards ${\bf n}=(1,0,0)$. 
(b) The profiles of the temperature field $\mathcal{T}(x)$ at  $t=2.5$ and  $t=100$ de facto coincide.
$\mathcal{T}(x)=0$ corresponds to the (lower) critical temperature $T_c$. Time is given in units of $t_0=3.6\times 10^{-8}$s.}
\label{fig:fig10}
\end{figure}

Figure~\ref{fig:fig10}(a)  shows a  snapshot of the temperature field in the $x$-$y$ and $x$-$z$ planes passing through the particle center (see Fig.~\ref{fig:fig1}). 
The numerical data correspond to the time  $t=100$ after  switching on the illumination.
In Fig.~\ref{fig:fig10}(b)
we plot the profiles of the temperature field  along the straight line parallel to ${\bf n}$ and running through the particle center. 
The two curves corresponding to early ($t=2.5$) and late ($t=100$) times are indistinguishable, which demonstrates that after the quench the steady state is achieved very quickly. 
The positions $x/R = 1$ and $x/R = -1$ correspond to the capped and  uncapped  surfaces, respectively. 
Since the capped side of the particle surface is heated more
strongly than the uncapped side, the temperature field has a peak at $x/R = 1$. The temperature is higher than  the lower  critical temperature ($\mathcal{T}=0$) for distances  up to about $x/R = 2$.
For larger distances it  attains  the background temperature $\mathcal{T}_i = -1$.  On the uncapped side, the temperature is higher  than the critical temperature up to about $x/R = -1$. 
We find that the temperature field is almost independent of the wettability parameters $h_{s,c}$ and $h_{s,l}$.

\begin{figure}[thpb!]
\includegraphics*[width=0.8\textwidth]{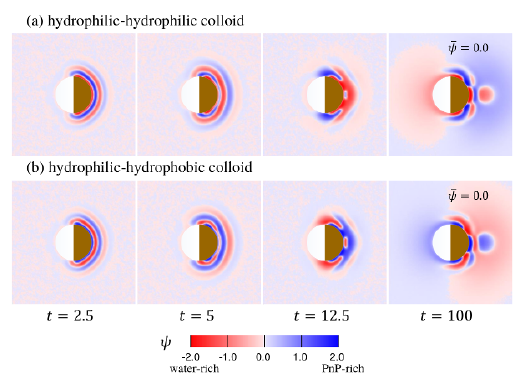}
\caption{The pattern evolution of the concentration field $\psi$ of a  mixture with $\bar{\psi}=0$  (see Eq.~(\ref{eq:aver_psi})) outside  a Janus particle and conserved as function of time, which has (a) a strongly \textit{hydrophilic} cap ($h_{s,c}=-2.0$) and a weakly \textit{hydrophilic} tail ($h_{s,l}=-0.2$); (b) a strongly \textit{hydrophobic} cap ($h_{s,c}=2.0$) and a weakly \textit{hydrophilic} tail ($h_{s,l}=-0.2$)
 As in Figs. 5 and 7, the brownish color indicates that the right hemisphere is capped. We do not specify  wettablity of the cap via  a color code.
}
\label{fig:fig11}
\end{figure}

\begin{center}
 \begin{figure}[tpbh!]\label{fig:fig12}
\includegraphics*[width=0.8\textwidth]{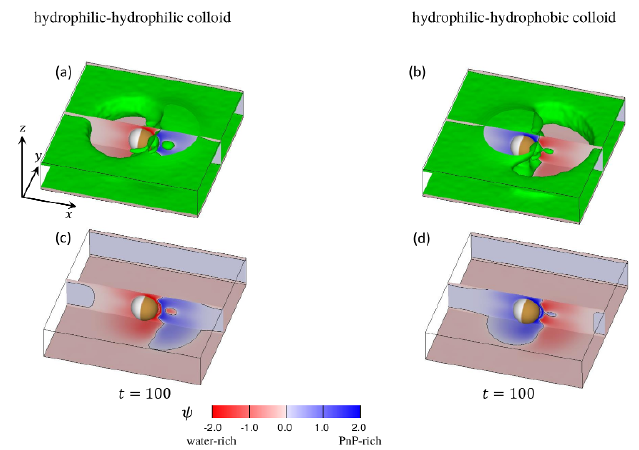}
\begin{spacing}{1.0}
\caption{Three-dimensional snapshots of $\psi$ reveal the complexity of the coarsening
patterns around a Janus particle at $t=100$ after illumination. The mixture is
at its critical concentration, i.e., $\bar\psi =0$. The panels (a) and (b) correspond
to a strongly \textit{hydrophilic} cap ($h_{s,c} =-2.0$) and a weakly \textit{hydrophilic} tail ($h_{s,l} = 0.2$) whereas the panels (b) and (d) correspond to a strongly
\textit{hydrophobic} cap ($h_{s,c}=2.0$) and a weakly \textit{hydrophilic} tail ($h_{s,l}=-0.2$). The
cross sections of the concentration field at the $x - z$ plane passing through the
particle center are shown in red-blue color, in line with Fig. 11 for $t=100$.
In front of and behind this $x-z$ plane, in (a) and (b) we show (in green) the
isosurfaces $\psi =0$. They consist of a   pair of
horizontal, flat sheets (one at the top and one at the bottom) connected by curved vertical surfaces. The sheets are slightly
separated from the weakly hydrophilic
confining walls at the bottom and at the top (with the top wall not shown) by a
thin film rich in water (marked reddishly). The connecting  surfaces  comprise the
external borders of two   nonspherical blobs:  one  rich in water (reddish areas to the
left (right) of the Janus colloid in (a) (in (b)) and the second (smaller)  rich in PnP (blueish
areas to the right (left) of the colloid in (a) (in (b)). The wavy  interface $\psi = 0$ between these two blobs  lies slightly to the right of the capped hemisphere.    Close to the hot
capped hemisphere (marked in brown) one can see strong fluctuation of the
concentration field, in particular, a droplet (in green) of the water-rich
phase in (a) and the PnP-rich phase in (b). In the panels (c) and (d) the
isosurfaces $\psi = 0$ are removed and one can see the concentration field at the
bottom wall. The interface positions $\psi = 0$  on the bottom and on the $x-z$
plane are indicated by black curves. Periodic boundary conditions are applied
along the $x$-axis.} 
\end{spacing}
\end{figure}
\end{center}

The time evolution of the concentration field coupled to the solvent flow  is depicted in Figs.~\ref{fig:fig11}(a) and (b) for the hydrophilic-hydrophilic and hydrophilic-hydrophobic particle, respectively. The snapshots  are taken at the same cross-sectional $x$-$y$ plane as the temperature field shown in Fig.~\ref{fig:fig10}(a). 
Only at very early  times  after the thermal  quench one can see the formation of layers which propagate from the surface of the Janus particle into the bulk. Later, the evolution of the coarsening patterns near the capped side of the colloid  becomes more complex  (see the movies in the SM). 
Already at  $t=12.5$  the layer structure becomes unstable. According to Fig.~\ref{fig:fig11}(a) the second rich-in-water layer (i.e., the overall third layer counted from the colloid surface) deforms, leading to the formation of  two bridges connecting this layer to the first one which is adjacent to the surface. In (b) the bridge formation occurs, too, but now formed by the PnP-rich layer.
Then the layers coalesce into a
 single droplet adjacent to  the capped side of the Janus particle  and surrounded by a depletion region (not shown here, see the movies in the SM). This configuration is also unstable.
In the course of time, new types of fluctuations of the concentration field appear -- mostly at the outermost part  of the phase-separating region but  also  close to the surface of the colloid along the border line between capped and uncapped hemispheres.
The pattern emerging  at $t=100$  is already quite complex such as the nucleation of a small droplet further away from the surface. 
Remarkably, at this late stage of the time evolution, the concentration pattern near the  left side of the colloid is determined entirely by the capped side.
Figure~\ref{fig:fig12} presents this pattern in the $x$-$z$  plane passing through the center of a colloid. 
One can clearly see the phase separation of the binary solvent  in the vicinity of the colloid. In the case of the hydrophilic-hydrophilic colloid, a
 region rich in the phase with $\psi <0$ (water-rich, red) has a shape of  a droplet which surrounds   the uncapped part of the colloid. Near the surface of the capped hemisphere,  both phases are present. Further away from it, only the phase with $\psi > 0$ (PnP-rich, blue) prevails.
 The coarsening pattern around the hydrophilic-hydrophobic colloid  is similar, only with the phase with $\psi >0$ being replaced by the phase with  $\psi <0$. 
  This demonstrates that within the hydrodynamic approach used here the wettability of the uncapped hemisphere is not important, provided the corresponding surface fields are weak.
In the present examples the wettabilities of the capped hemispheres  of both types of Janus particles are opposite to
each other,  but their strengths are the same.
In this case, the figures showing the coarsening patterns look the same but with a different coloring.

\begin{figure}[htpb]
\includegraphics[width=0.9\textwidth]{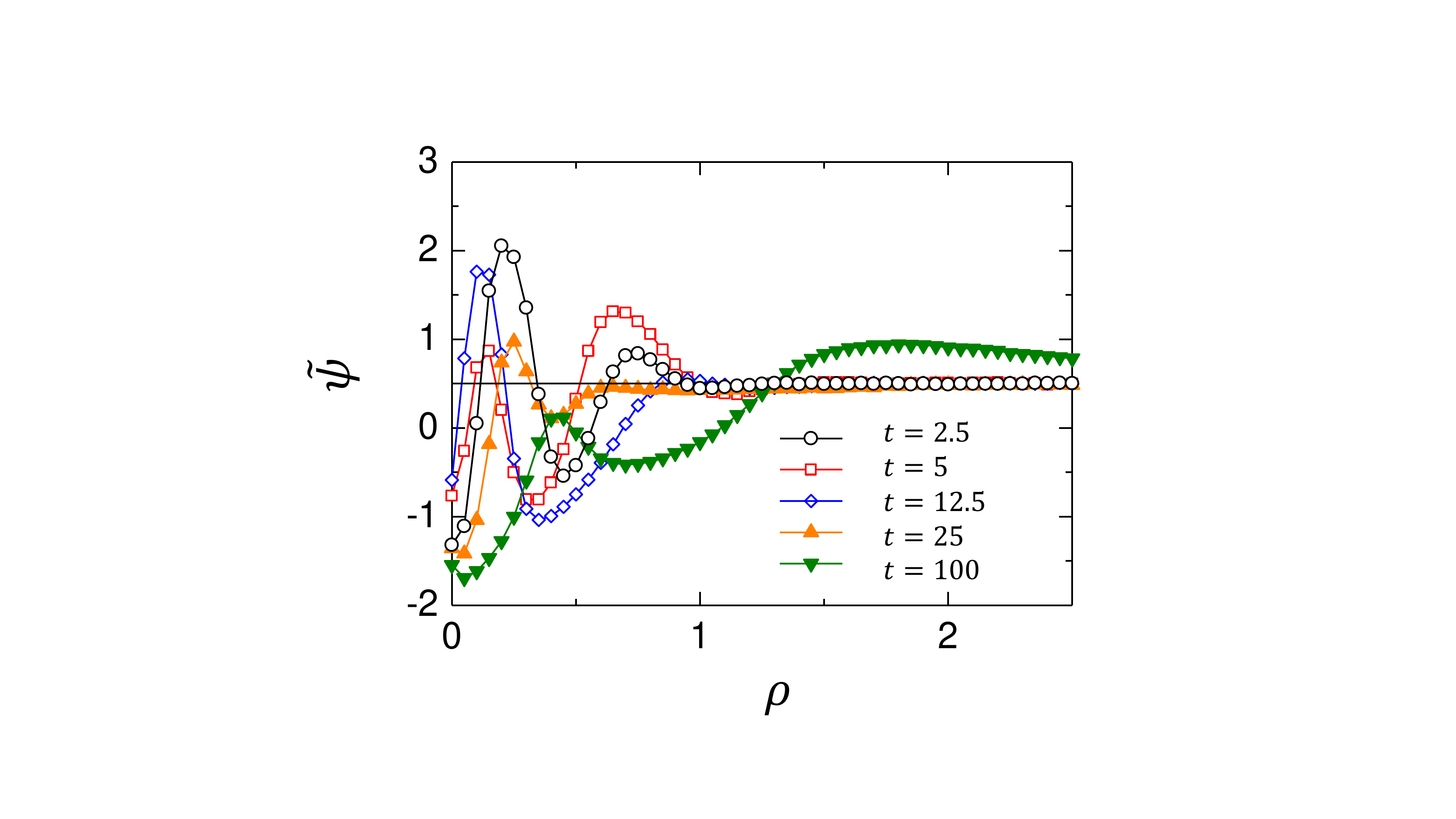}
\caption{Evolution of the rescaled concentration profiles  $\tilde\psi(\vec r,t)= (\psi(\vec r,t)+1)/2$ (not averaged) along the $x$-axis around a \textit{hydrophilic-hydrophilic} Janus colloid as a function of the radial distance $\rho = (r-R)/R$ from the surface of the colloid,
corresponding to the snapshots shown in Fig.~\ref{fig:fig12}(a).
The time is given in units of $t_0=10^{-6}$s. The horizontal black line indicates the initial bulk value  $\tilde\psi(\rho \to\infty)= 0.5$. }
\label{fig:fig13}
\end{figure}

\begin{figure}[htpb]
\includegraphics[width=0.9\textwidth]{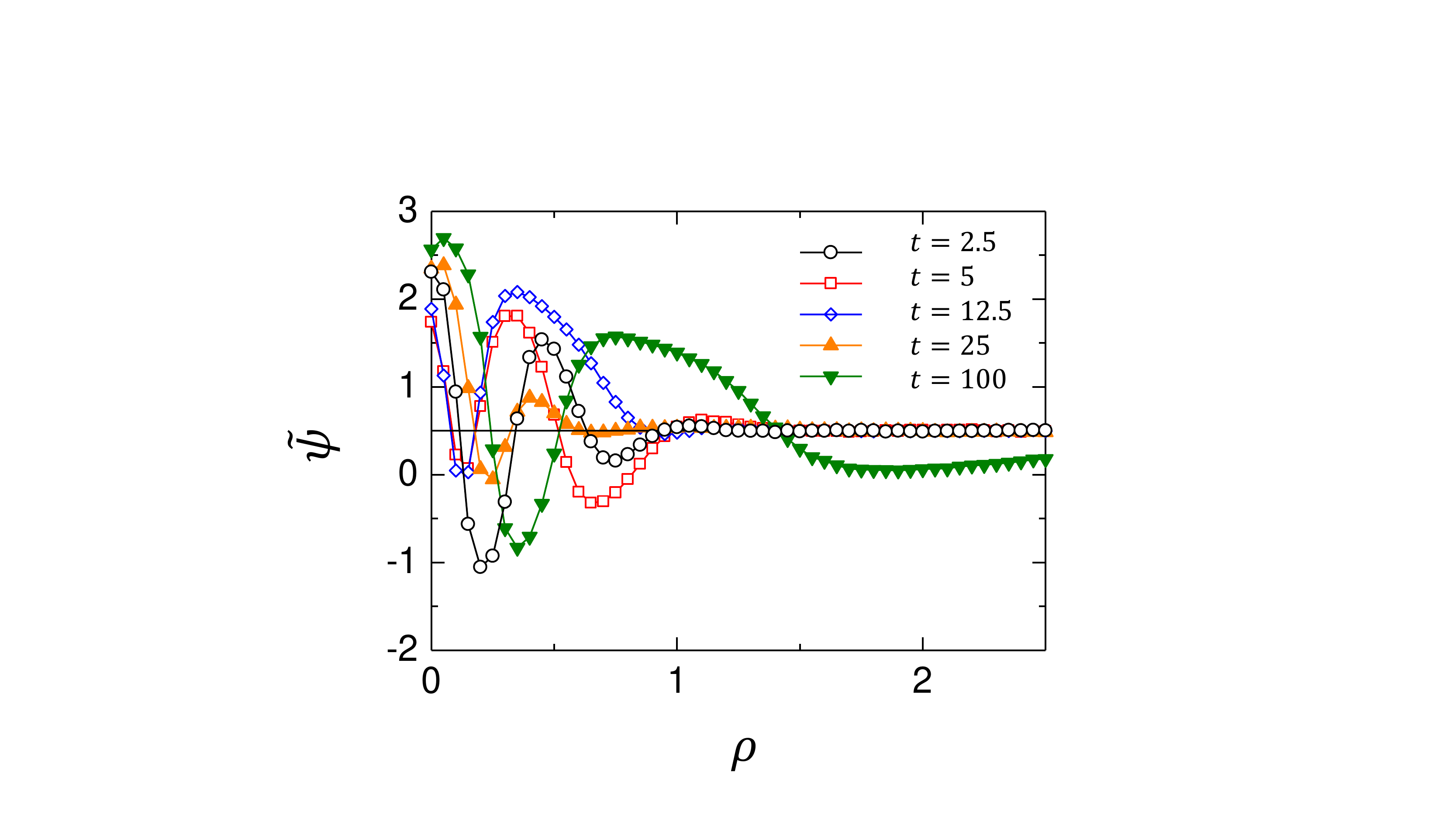}
\caption{Evolution of the rescaled concentration profiles $\tilde\psi(\vec r,t)= (\psi(\vec r,t)+1)/2$ (not averaged) along the $x$-axis around a \textit{hydrophilic-hydrophobic} Janus colloid as function of the radial distance $\rho = (r-R)/R$ from the surface of the colloid,
corresponding to snapshots like the ones shown in Fig.~\ref{fig:fig12}(b).
The time is given in units of $t_0=10^{-6}$s. The horizontal black line indicates the initial bulk value  $\tilde\psi(\rho \to\infty)= 0.5$. }
\label{fig:fig14}
\end{figure}

Profiles of the rescaled concentration field  $\tilde\psi(\vec r,t)= (\psi(\vec r,t)+1)/2$   along the $x$-axis as a function of the radial distance $\rho = (r-R)/R$ from the surface of Janus particles,  are shown in Figs. \ref{fig:fig13} and \ref{fig:fig14}.
They  correspond to the snapshots shown in  Figs.~\ref{fig:fig12}(a) and (b), respectively.
Only  at early times ($t=2.5$) the profiles look similar to those obtained from the diffusive model as shown in Fig.~\ref{fig:fig9}. For later times, strong concentration fluctuations deform the the radial  structure  of the profiles.
The time  evolution of the concentration profiles   near the left side of  the Janus colloids reveals the role of the capped hot side 
(see Figs.~3 and 4 in SM). The left side is  always hydrophilic and 
at early times one can see  a very small water-rich  layer near it. If the cap is hydrophilic, this layer becomes much thicker at late times. 
If the  cap is hydrophobic, at later times this layer is smeared out by a PnP-rich concentration wave traveling from the right side of the Janus colloid. 

The above results raise the issue concerning the origin of the strong  concentration fluctuations  observed within the hydrodynamic approach.
It is well established that  thermal fluctuations in liquids in the presence of stationary temperature gradients are anomalously large and very long-ranged~\cite{Sengers}.
They occur as a result of  a coupling between temperature and velocity fluctuations. 
It is to be expected that in liquid mixtures a 
temperature gradient induces long-ranged concentration
fluctuations via the Soret effect. Moreover, these nonequilibrium fluctuations have to  be coupled to the critical concentration fluctuations. Indeed, we observe that the concentration fluctuations appear mostly in that region where
the temperature is critical or close to its  critical value.

\subsubsection*{B.4 Comparison with experiment}\label{arakimodel:4}

At very short times after the temperature quench, the results  of the hydrodynamic model agree with those obtained from  the purely diffusive model as well as with experimental observations: compare, e.g., the layer structures 
at $t_{exp}=0.2$s and  $t_{exp}=0.4$ in Figs.~\ref{fig:fig2}(b) and (c), at  $t=$ 0.8 and $t=2$ in Figs.~\ref{fig:fig5} and ~\ref{fig:fig6},  and at $t=2.5$ and $t=5$ in Figs.~\ref{fig:fig11}(a) and (b). 
In the hydrodynamic model, however, the lifetime of the composition waves is much shorter than in a purely diffusive model. The same shortness was found  for the composition wave near a planar wall, which forms after a 
temperature quench in the bulk~\cite{EurophysLett.51.154}.
This very short lifetime prohibited a reliable determination  of  the speed of the transient layers. 
Very quickly  strong fluctuations of the concentration field lead to a  destabilization of the layer structure so that we could not observe the  continuous growth of a droplet.
These fluctuations may explain the experimental results obtained for a Janus particle with a hydrophilic or  hydrophobic cap,
for which at $t_{exp}=0.8$s  or  at $t_{exp}=0.53$s, respectively,  the layer structure fluctuates and even gets destroyed.
Instead, one can see the formation of  a flower-like pattern around the hot hydrophobic  cap of the colloid.

\section{Conclusions}\label{concl}

The  important conclusion from our study is that at early times after applying laser illumination 
to the sample, the coarsening dynamics around Janus colloids is dominated by diffusion.
It is characterized by the formation of transient layers propagating with constant speed from the surface of the Janus particle into the bulk.
In the case of an immobilized particle, after a transient time the coarsening process is dominated by hydrodynamic effects with the patterns continually varying in time.

\begin{figure}[htpb]
\includegraphics[width=0.8\textwidth]{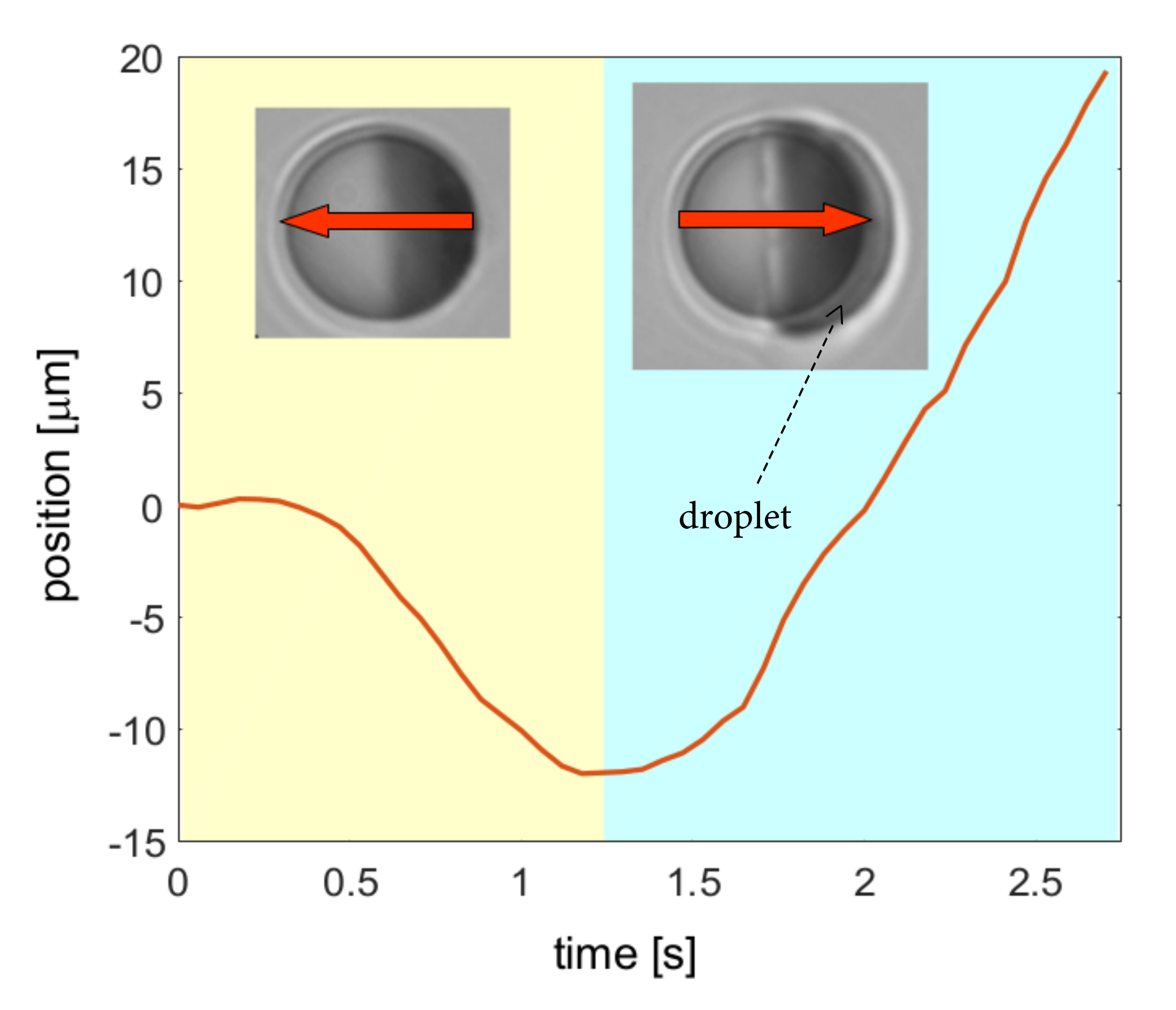}
\caption{Position of an experimentally moving colloid as function of time after an inverse temperature quench. 
This illustrates the 
self-propulsion of the particle during the transient formation of a droplet. }
\label{fig:fig15}
\end{figure}

These spatio-temporal patterns are different if the particle size is small enough so that it can move 
between the two parallel confining walls ($h>2R$, Fig.\ref{fig:fig1}(a)). The transient layering   
precedes the formation of a droplet, which characterizes the stationary state of the fluid surrounding a moving colloid.
In \cref{fig:fig15}, we show our data for the position of the center of a mobile colloid during the transient demixing (yellow area) and 
in the steady state (blue area). The arrows indicate the corresponding directions of motion. 
For these measurements we have used 2,6-lutidine-water as a binary solvent, and the particle has a hydrophobic golden cap 
on the right hemisphere. The particle can move through the fluid 
because it is slightly smaller in diameter $2R$ than the thickness $h$ of the sample cell. The laser illumination is turned on at the beginning of the video (see SM). During the first 
0.4 seconds, the colloid remains at rest, but thereafter it starts to move with rather constant speed 
in the direction opposite to the golden cap (right hemisphere), while the transient layers develop. Finally, after 1.2 seconds it reverses 
its direction of motion, once the droplet has almost reached its final size. The speed during the initial 
transient is about 8 microns per second, so that the effect of the reversal of the direction is rather large. 

Because the colloid remains at rest during the first 0.4 seconds, we can employ our  simulation models developed for  {\it fixed} particle
in order to  calculate the body force due to the concentration flux  acting on a fixed colloid immediately  after an inverse temperature quench.
We have used a purely diffusive approach, which we found to be an appropriate description of the early time  dynamics. 
The force is calculated as 
\begin{equation}
 \label{eq:body_force}
\vec F(\vec r,t) = \int_{\mathscr{V_C}} d^3 r \psi(\vec r,t)\nabla \mu(\vec r,t), 
\end{equation}
where the chemical potential $\mu(\vec r,t) = \nabla (\delta {\mathcal F[\psi]}/\delta \psi(\vec r,t))$ follows from the free energy functional of the CHC theory (see Appendix~\ref{app:CHC}).
$\mathscr{V_C}$ refers to the volume of the colloid.

\begin{figure}[htpb]
\includegraphics[width=0.8\textwidth]{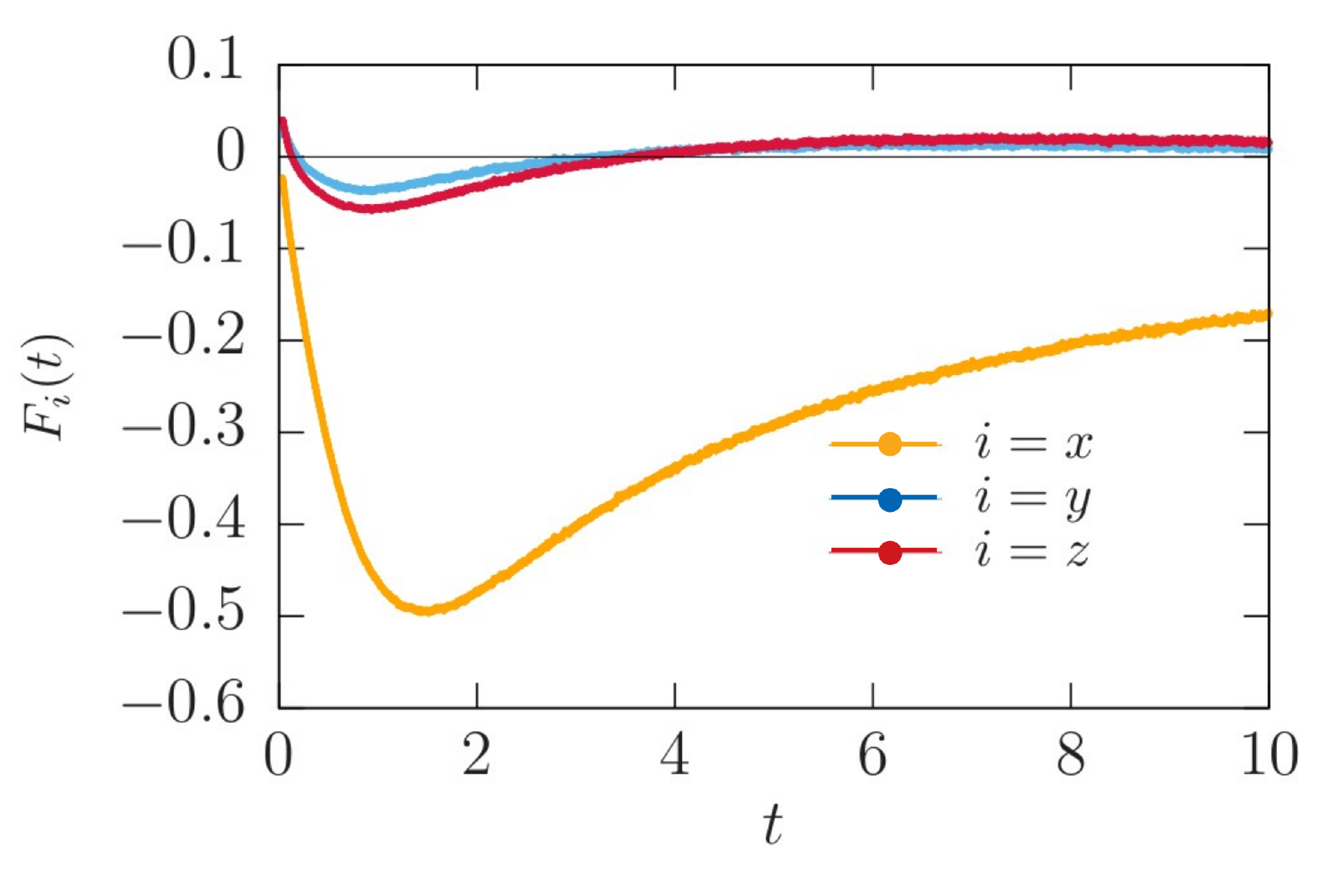}
\caption{Time dependence of the  $x, y$, and $z$ components  of the theoretical body force $\vec F (t)$  (Eq.~(\ref{eq:body_force})) acting on the Janus particle  after the temperature quench. Time is given in units of $t_0=10^{-6}$s.
}
\label{fig:fig16}
\end{figure}

In \cref{fig:fig16}, we present numerical results for the three Cartesian components  of  $\vec F(\vec r,t)$ acting on  a hydrophilic-hydrophobic Janus colloid after an inverse temperature quench was applied at $t=0$.
The results correspond to $L_x=L_y=L_z=100$, $R=5$, $\mathcal{T}_i=-1$, $\mathcal{T}_1=8$, $s=0.001$, $\mathcal{D}=100$, $\mathcal{D}_c=300$, $\alpha_{s,l}=\alpha_{s,c}=0.5$, $h_{s,l}=2$, and $h_{s,c}=-2$. 
One can see that after some transient  the magnitude of the $x$-component of the force  increases,  and that beyond a certain time it starts to decrease. The force vanishes at about $t\simeq 100$ when the steady state is achieved, i.e., after 
the local
demixing has been completed, and a droplet, rich in the species preferred by the hot hemisphere of
the Janus particle, has been formed.
The strength of the $x$-component is much bigger than the strengths of the $y$- and $z$-components. 
This is the reason for the onset of the  particle motion in $x$- direction. 
Once the colloid is set into motion, our simulation model does no longer apply. The constant speed observed between 0.5 and 1.0 seconds (Fig. 15) indicates
that the total force acting on the particle  is zero,  although  the coarsening process has  not yet attained its steady state.
It is plausible that at these early times the body force $\vec F(\vec r,t)$ is vanishingly small and concurrently,  hydrodynamic forces are still absent. The initial motion of a colloid is stopped by fluid resistance. But in the meantime hydrodynamic effects become relevant for the coarsening 
process, giving rise to a propulsion force which changes the direction of motion of the colloid. 

Finally, an interesting conclusion can be drawn from those results, which are obtained by employing 
a hydrodynamic model. These results clearly demonstrate the high relevance of nonequilibrium concentration fluctuations, induced by 
temperature gradients, for the coarsening dynamics.
It would be rewarding to study in more detail how   these nonequilibrium concentration fluctuations couple to the critical fluctuations in the vicinity of the demixing critical point of the binary liquid mixture.

\section{Conflicts of interest}
There are no conflicts of interest to declare.

\textbf{Acknowledgments:} The work by AM has been supported by the Polish National Science Center 
(Harmonia Grant No. 2015/18/M/ST3/00403). TA acknowledges the support from JSPS KAKENHI Grant
Number JP17K05612, and JST CREST Grant No. JPMJCR1424, Japan. 
J.R.G.-S. acknowledges support from DGAPA-UNAM PAPIIT Grant No. IA103320.

 \newpage
\clearpage

\clearpage

\clearpage

\clearpage

\clearpage

\clearpage

\clearpage

\clearpage

\clearpage

\clearpage

\clearpage

\clearpage

\clearpage

\clearpage

\clearpage

\clearpage

\newpage
\appendix
\section{Diffusive model}
\label{app:CHC}

 Within the CHC theory the free energy functional of the 
solvent is given by
\begin{equation}\label{CHC1} 
\frac{\mathcal{F}}{k_BT_c} = \int_{} \frac{d^{3}r}{\mathrm{v}_0^3}\Bigl[\frac{1}{2}c\bigl(\nabla \psi(\vec r)\bigr)^{2} - \frac{1}{2}{a} \psi(\vec r)^{2}
+ \frac{1}{4} u  \psi(\vec r)^{4}\Bigr],
\end{equation}
where $c$, $a$, and $u$ are phenomenological parameters with $a~=\mathcal{A} (T-T_c)/T_c \equiv \mathcal{A} \tau$ and 
$T_c$ as the demixing critical temperature for a binary liquid mixture. Accordingly, $a$ is 
related to the correlation length of the fluid which diverges upon approaching $T_c$. Here we consider a lower critical point so that $a<0$ for the mixed phase.
The  length scale  $\mathrm{v}_0$ is taken to be 
 a microscopic length scale (e.g., the diameter of the molecules of the solvent). 
For a conserved OP field its time evolution is related to the 
free energy as
\begin{equation}\label{CHC2} 
\frac{\partial \psi(\vec r,t)}{\partial t} = -\nabla \cdot {\vec j(\vec r, t)} = - \nabla \cdot  [-M \nabla \mu(\vec r, t)]  =
M \nabla \frac{\delta {\mathcal F[\psi]}}{\delta \psi(\vec r,t)},
\end{equation}
$\vec j$ and $M$ being the concentration current and the mobility, respectively. The concentration current is proportional to the gradient of the local chemical
potential  $\mu(\vec r, t)$.  This leads to the 
CHC equation at \textit{constant} temperature $T$:
\begin{equation}\label{CHC3} 
\frac{\partial \psi(\vec r,t)}{\partial t} = \frac{M}{\mathrm{v}_0^3} k_B T_c \nabla ^2 \Big (-a\psi(\vec r,t) + 
u\psi^3(\vec r,t) - c\nabla^2 \psi(\vec r,t) \Big)+ \zeta(\vec r,t);
\end{equation}
$\zeta(\vec r,t)$ is a Gaussian random {white  noise which describes   the thermal fluctuations: 
\begin{equation}\label{CHC4} 
\langle \zeta(\vec r,t) ~\zeta(\vec r', t')\rangle= -2(M/\mathrm{v}_0^3) k_B T_c \nabla^2 \delta (\vec r-\vec r') \delta (t -t').
\end{equation}
Here we have assumed that local equilibrium prevails so that the noise obeys
the fluctuation-dissipation theorem.

In order to extend the CHC model to the the present colloidal system, in which the temperature field   depends both on space and time,
the coefficient $a$ in \cref{CHC3} is replaced by $\mathcal{T}(\vec r)={\mathcal A} (T(\vec r)-T_c)/T_c)$ where 
$\mathcal A$ is a dimensionless constant.  By using suitable substitutions, \cref{CHC3} takes on a 
modified dimensionless form:
\begin{equation}\label{CHC5} 
\frac{\partial \psi(\vec r,t)}{\partial t} = \nabla ^2 \Big (-\frac{\mathcal{T}(\vec r,t)}{|\mathcal{T}_1|} \psi(\vec r,t) + 
\psi^3(\vec r,t) - C\nabla^2 \psi(\vec r,t) \Big)+ \zeta(\vec r,t),
\end{equation}
where
\begin{subequations}\label{CHC6} 
\begin{align}
\psi \to \psi_0\psi, \quad \psi_0 & = \sqrt{|\mathcal{T}_1| /u},\\
\vec r \to r_0 \vec r,\quad r_0 &  = \sqrt{\frac{2}{c_0}}\xi_-(T_1),\\
     t \to  t_0t, \quad  t_0 &  = \frac{2}{c_0}\xi^2_-(T_1)/(D_m(T_c)|\mathcal{T}_1|).\\
\end{align}
\end{subequations}
$\mathcal{T}_1$ is the reduced quench temperature of the cap,  $\psi_0$ is the absolute value of the mean-field bulk OP at $T=T_1$, and $C=c/c_0$ is  a dimensionless parameter.
In Refs.~\cite{Roy-et:2018,Roy-et:2018a,Araki:2019}, which are concerned with  shallow  quenches, this parameter was set to 1, i.e., the rescaling factor $c_0$ was equal to $c$. In the present study we assume $C=4$, which is  appropriate for deep
quenches, as considered  here, for which the bulk correlation length
$\xi_-(T_1)$ at $T_1$ is comparable to the  size of the mesh in the numerical calculations. 
We assume that the strength  $\zeta_0(\vec r,t)$ of the dimensionless noise $\zeta(\vec r,t)$ is  uniform in space and is  expressed in units of  $\eta_0 = \psi_0/t_0$.
$D_m(T_c)=(M/\mathrm{v}_0^3) k_B T_c$ is the interdiffusion constant of a binary solvent at the critical temperature.

We recall that for the system with a homogeneous temperature $T$, the mean field bulk correlation length (above $T_c$ in the system with a lower critical point)
 is given by $\xi_{-}(T)=\xi_0^-|\tau|^{-1/2}$,  where the amplitude 
 $\xi_0^-=\sqrt{c/{\mathcal A}}$ follows from the free energy  given by Eq.~(\ref{CHC1}).  
Using the relations  $\mathcal{T}(\vec r)={\mathcal A} (T(\vec r)-T_c)/T_c)$ and $\xi_- \simeq \xi_0^-|T(\vec r)-T_c)/T_c)|^{-\nu}$ with the critical exponent $\nu \simeq 0.63$ \cite{Pelisseto} 
of the three-dimensional Ising model and assuming typical values for the bulk correlation
length amplitude for binary liquid mixtures,  $\xi_0^-\simeq 0.1$nm \cite{mirzaev2006,Gulari}, $c\simeq 1.85$nm$^2$ \cite{Roy-et:2018},  and for the amplitude ${\cal A}=c/(\xi_0^-)^2 \simeq - 46.3$, we obtain $\mathcal{T}_1 =6$ for the quench temperature and  $r_0 \simeq 0.35$nm for the length unit. Approximating  $D_m(T_c)$ by a   value of $D_m$ at a rather small 
reduced temperature $|(T-T_c)|/T_c =10^{-6}$, which typically is of the order 
 $10^{-14}$m$^2/$s \cite{mirzaev2006}, renders $t_0\simeq 10^{-7}$s for the time unit.

The 
generic preference of the colloidal surface for one of the two components of the binary mixture is taking account by  considering  a 
surface energy contribution $\frac{1}{2} \alpha_s \int_{\mathscr S} { \psi}^2 dS - 
h_s \int_{\mathscr S} { \psi} dS$ 
which has to be added  to the free energy functional in Eq.~(\ref{CHC1}) \cite{diehl1997}. Here, $\mathscr S$ refers to the surface of the colloid, $\alpha_s$ is a surface enhancement parameter, and $h_s$ is a 
symmetry breaking surface field. Upon the substitutions $\alpha_s \to  (c/|\mathcal{T}_1|)^{-1/2}\alpha_s$
and $h_s \to  \Big((|\mathcal{T}_1|/u)^{1/2}/(c/|\mathcal{T}_1|)^{-1/2}\Big) h_s$, this contribution  leads to  the dimensionless static 
 Robin b.c. as given by Eq.~(\ref{bc2}).
 
 \newpage

\newpage
\section{Supplementary Material}
\par
\hspace{0.2cm}\textbf{Diffusive dynamics} 

Additional information about the formation of the layers can be gained by inspecting 
the vector snapshots of the OP flux which is defined as the spatial gradient of $\psi(\vec r,t)$ normalized to one. 
The results shown in Fig.~\ref{Sfig:1} are also depth-averaged  but not averaged over initial conditions. 
One can see the white and red ``lines'' around the colloid, which correspond 
to the minima (water-rich layers) and maxima (water-poor depletion layers) of the OP profile, respectively. 
Following the inverse thermal quench, at first the number of lines increases indicating the formation of surface layers. 
At later times, the number starts to decrease until at $t=1000$ one observes only a single depletion layer. 
At $t=1000$, away from the colloid the fluid is uniformly mixed as demonstrated by the red and white points. 

\begin{figure}[htbp]
\centering
\includegraphics[width=0.9\textwidth]{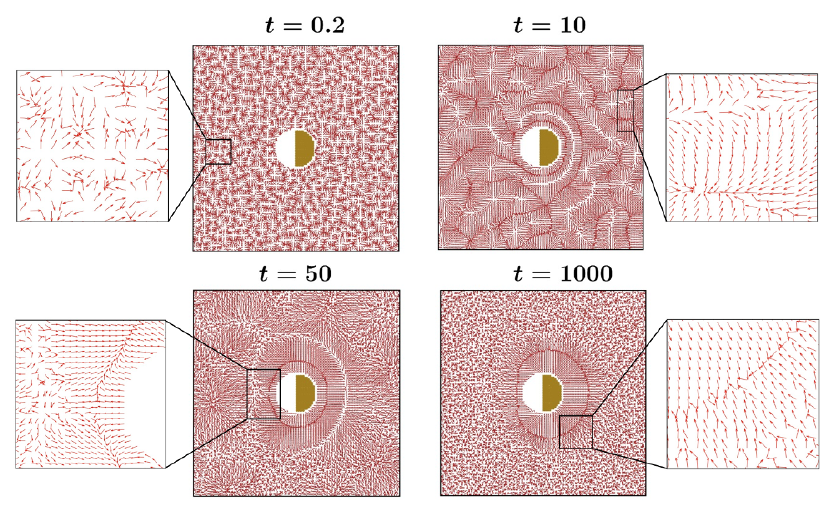}
\caption{Normalized order parameter flux $\nabla \psi(\mathbf{r},t)/|\nabla \psi(\mathbf{r},t)|$ 
around the \textit{hydrophilic-hydrophilic} Janus colloid studied in the main text. 
Time is given in units of $t_0=10^{-6}$s. The set of parameters is the same as for Fig.~5 in the main text. The flux of the OP is colored differently than $\psi(\mathbf{r},t)$  in Fig.~5 in the main text. }
\label{Sfig:1}
\end{figure}

Figure \ref{Sfig:2} elucidates the temporal evolution of the order parameter flux around a hydrophilic-hydrophobic
colloid. 

\begin{figure}[ht]
\centering
\includegraphics[width=0.9\textwidth]{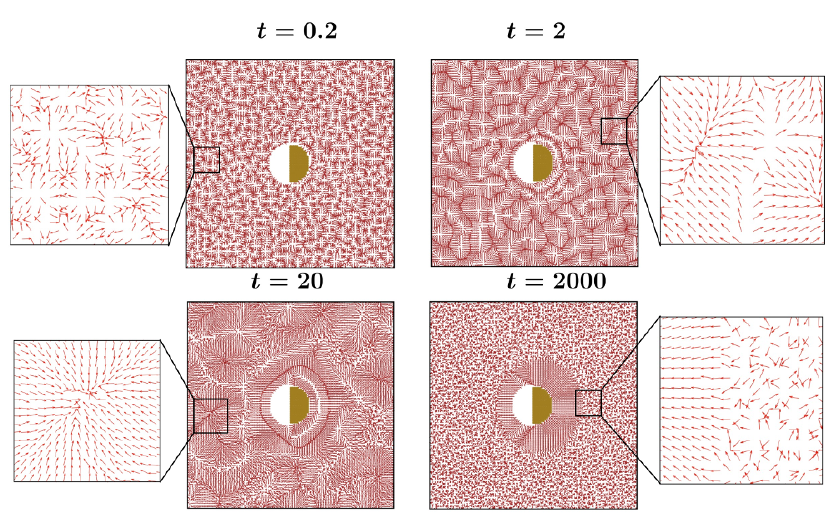}
\caption{Normalized order parameter flux $\nabla \psi(\mathbf{r},t)/|\nabla \psi(\mathbf{r},t)|$  around the \textit{hydrophilic-hydrophobic} Janus colloid studied in the main text. The temporal development is similar to the one for the hydrophilic-hydrophilic particle shown in Fig.~\ref{Sfig:1} of SM, but the steady state is different.
It reflects the formation of a droplet rich in water on the left side of the Janus particle, 
which is missing in Fig.1 in SM for $t=1000$.  Time is given in units of $t_0=10^{-6}$s. The set of parameters is the same as for Fig.~7 in the main text. The flux of the OP is colored differently than $\psi(\mathbf{r},t)$  in Fig.~7 in the main text.
}
\label{Sfig:2}
\end{figure} 

\par
\hspace{0.2cm}\textbf{Hydrodynamic approach} 

Figures~ \ref{Sfig:3} and  \ref{Sfig:4} show  the  concentration profiles along the $x$-axis for the hydrophilic-hydrophilic and 
the hydrophilic-hydrophobic Janus colloid studied in the main text, obtained within the hydrodynamic approach.
The centre of the colloid is located at $x=0$. 
At early times ($t<25$), one can see small dips around $x/R=-1$. These dips are due to the
weakly hydrophilic wetting condition on the left side of the Janus colloid.
In the late stage, this dip is smeared out
by the concentration wave arriving from the right side of the Janus colloid.

\begin{figure}[ht]
\includegraphics[width=0.9\textwidth]{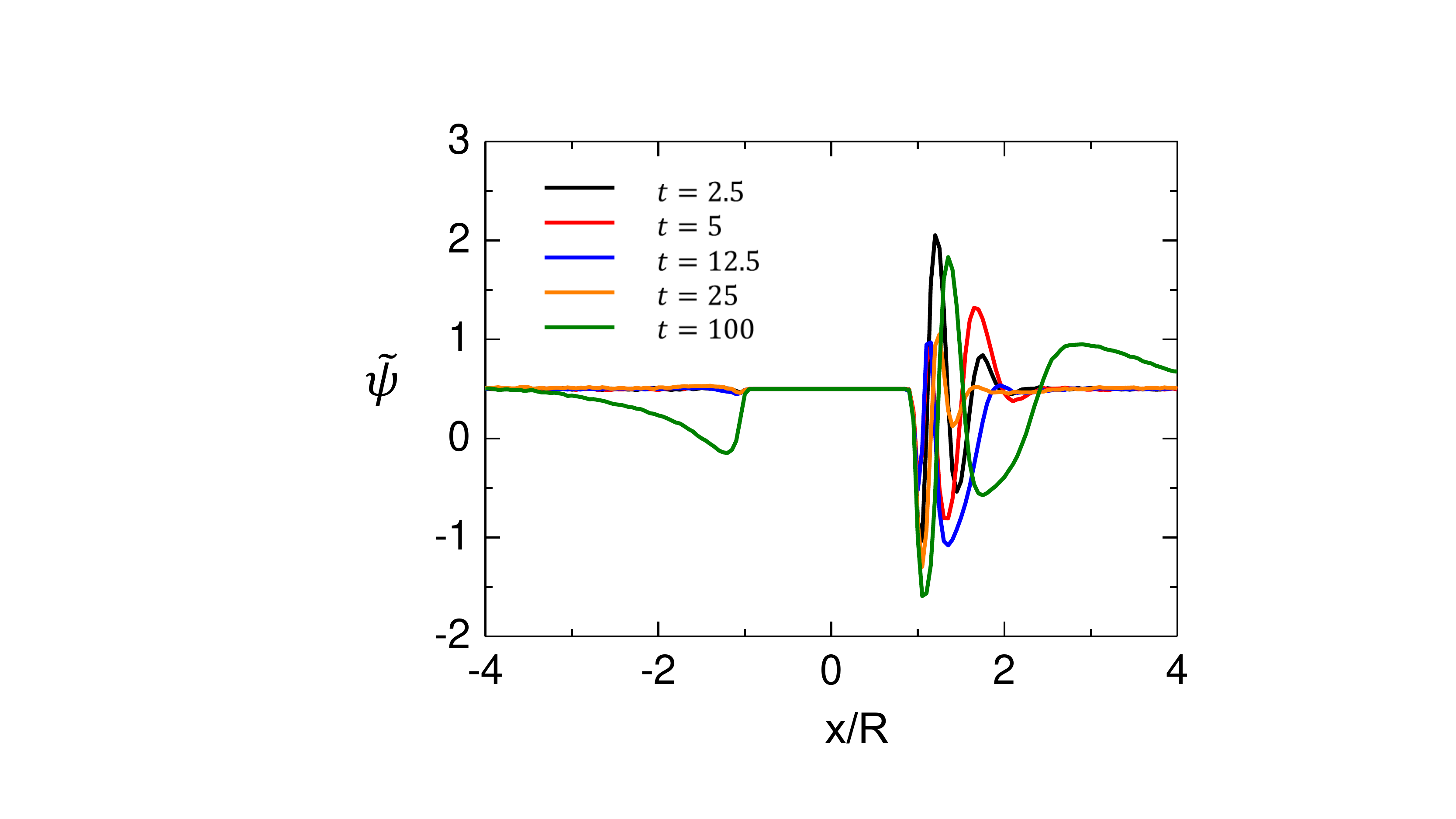}
\caption{Evolution of the rescaled concentration profile $\tilde\psi(\textbf r,t)= (\psi(\textbf r,t)+1)/2$  along the $x$-axis around the 
\textit{hydrophilic-hydrophilic} Janus colloid studied in the main text as function of the distance $x/R$ from the center of the colloid,
corresponding to the snapshots shown in Fig.~11(a) in the main text. In Fig.~13  in  the main text we show these  profiles in  full detail and  in an expanded interval of $x$, corresponding to the right side of the Janus particle, i.e., for $x/R > 1$.  At short times the concentration profile near the left (uncapped) side of the Janus colloid is slightly negative. This is the case because the left hemisphere is weakly hydrophilic. At late times the profile on this side of the Janus particle is dominated by the concentration wave arriving from the heated right hemisphere. In the case of the hydrophilic right side, this concentration wave  drives the profile close to the left side  even more negative. 
The time is given in units of $t_0=10^{-6}$s. The initial bulk value  $\tilde\psi(x \to\infty)= 0.5$. }
\label{Sfig:3}
\end{figure}

\begin{figure}[ht]
\includegraphics[width=0.9\textwidth]{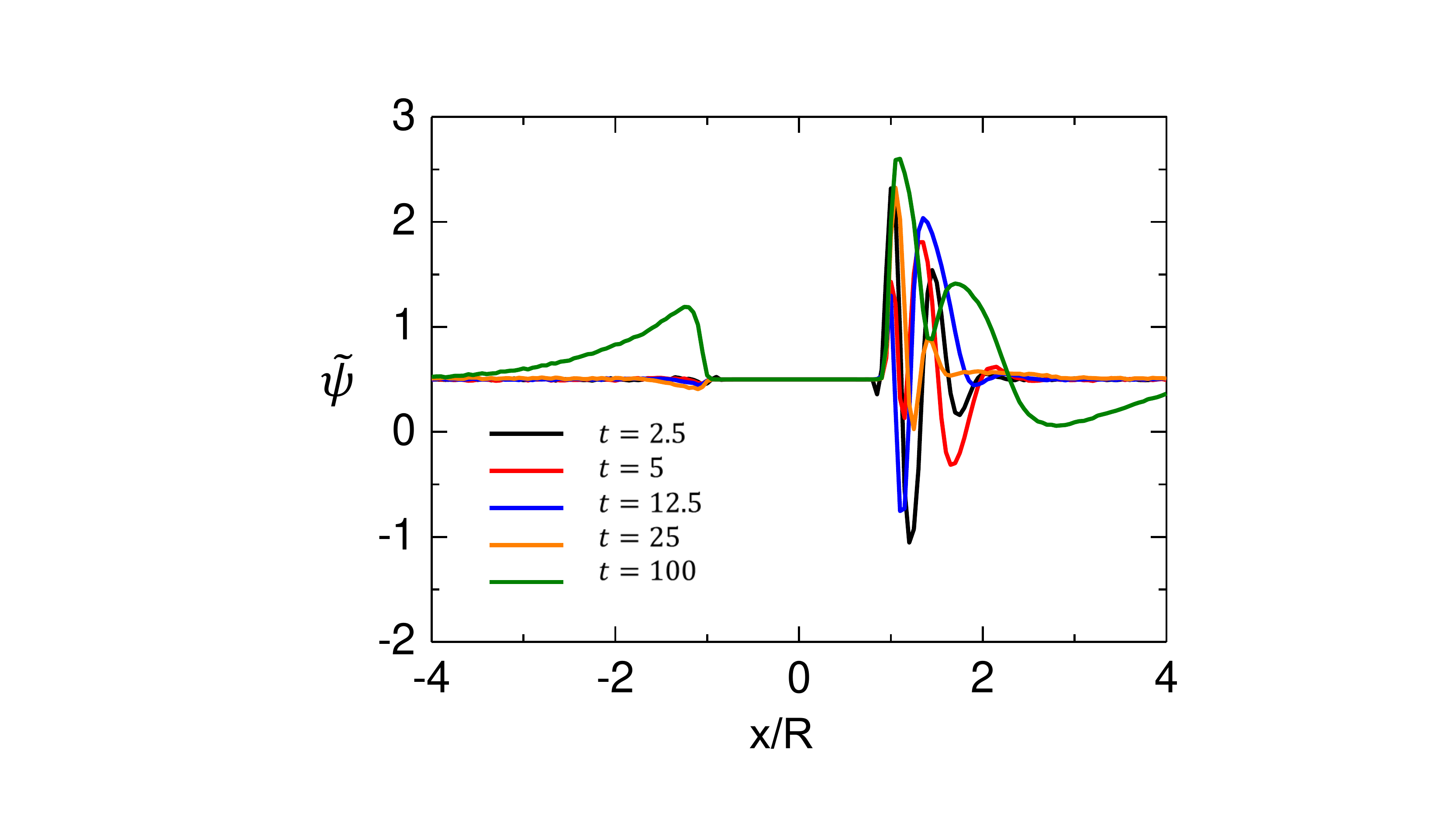}
\caption{Evolution of the rescaled concentration profile $\tilde\psi(\textbf r,t)= (\psi(\textbf r,t)+1)/2$ along the $x$-axis around the 
\textit{hydrophilic-hydrophobic} Janus colloid studied in the main text as function of the distance $x/R$ from the center of the colloid, corresponding to the snapshots shown in Fig.~11(b) in the main text. In Fig.~14 in  the main text we show these  profiles in full detail and in an expanded interval  of $x$, corresponding to the right side of the Janus particle, i.e., for $x/R > 1$. As in Fig.~3, at short times the 
concentration profile near the left (uncapped) side of the Janus colloid is slightly negative.  This is the case because the left hemisphere is weakly hydrophilic. At late times the profile at this side of the Janus particle is dominated by the concentration wave arriving 
from the heated right hemisphere. In the case of the hydrophobic right side, near the left side this concentration wave drives the profile   positive. 
The time is given in units of $t_0=10^{-6}$s. The initial bulk value  $\tilde\psi(x \to\infty)= 0.5$. }
\label{Sfig:4}
\end{figure}

 \end{document}